%% file: main.tex
\documentclass[twocolumn]{aastex701}

\graphicspath{{./}{figures/}}
\makeatletter
\def\input@path{{./}{chapters/}}
\makeatother

\newcommand{\um}{\,\mu\text{m}}

\usepackage{amsmath}

\begin{document}

\title{Empirical Near-Infrared Spectral Templates from SPHEREx:\\
Disentangling AGN and Host-Galaxy Emission}

\author[0000-0002-3742-6609]{Wenke Ren}
\affiliation{Shanghai Astronomical Observatory, Chinese Academy of Sciences, 80 Nandan Road, Shanghai 200030, People's Republic of China}
\email[show]{wenke.ren@shao.ac.cn}

\author[0000-0001-8416-7059]{Hengxiao Guo}
\affiliation{Shanghai Astronomical Observatory, Chinese Academy of Sciences, 80 Nandan Road, Shanghai 200030, People's Republic of China}
\email[show]{hengxiaoguo@gmail.com}

\author[0000-0002-9634-2923]{Zhenya Zheng}
\affiliation{Shanghai Astronomical Observatory, Chinese Academy of Sciences, 80 Nandan Road, Shanghai 200030, People's Republic of China}
\email{zhengzy@shao.ac.cn}

\author[0000-0002-4455-6946]{Minfeng Gu}
\affiliation{Shanghai Astronomical Observatory, Chinese Academy of Sciences, 80 Nandan Road, Shanghai 200030, People's Republic of China}
\email{gumf@shao.ac.cn}

\author[0000-0002-4521-6281]{Wenwen Zuo}
\affiliation{Shanghai Astronomical Observatory, Chinese Academy of Sciences, 80 Nandan Road, Shanghai 200030, People's Republic of China}
\email{wenwenzuo@shao.ac.cn}

\author[0000-0002-4419-6434]{Junxian Wang}
\affiliation{Department of Astronomy, University of Science and Technology of China, 96 Jinzhai Road, Hefei, Anhui 230026, People's Republic of China}
\affiliation{College of Physics, Guizhou University, Guiyang, Guizhou, 550025, People's Republic of China}
\email{jxw@ustc.edu.cn}

\author[0000-0003-3424-3230]{Weida Hu}
\affiliation{Shanghai Astronomical Observatory, Chinese Academy of Sciences, 80 Nandan Road, Shanghai 200030, People's Republic of China}
\email{weidahu@shao.ac.cn}

\author[0000-0001-8515-7338]{Zhen-Bo Su}
\affiliation{Department of Astronomy, University of Science and Technology of China, 96 Jinzhai Road, Hefei, Anhui 230026, People's Republic of China}
\affiliation{School of Astronomy and Space Science, University of Science and Technology of China, Hefei 230026, People's Republic of China}
\email{zbsu@mail.ustc.edu.cn}

\author[0009-0004-9950-9807]{Shi-Jiang Chen}
\affiliation{Department of Astronomy, University of Science and Technology of China, 96 Jinzhai Road, Hefei, Anhui 230026, People's Republic of China}
\email{JohnnyCsj666@gmail.com}

\author[0000-0003-3787-0790]{Lin Long}
\affiliation{Shanghai Astronomical Observatory, Chinese Academy of Sciences, 80 Nandan Road, Shanghai 200030, People's Republic of China}
\email{longlin@shao.ac.cn}

\author[0000-0002-3134-9526]{Jiaqi Lin}
\affiliation{Shanghai Astronomical Observatory, Chinese Academy of Sciences, 80 Nandan Road, Shanghai 200030, People's Republic of China}
\email{linjiaqi@shao.ac.cn}

\author[0009-0002-5955-4932]{Chi-Zhuo Wang}
\affiliation{Department of Astronomy, School of Physics, Peking University, Beijing 100871, People's Republic of China}
\affiliation{Kavli Institute for Astronomy and Astrophysics, Peking University, Beijing 100871, People's Republic of China}
\email{chizhuowang@stu.pku.edu.cn}

\begin{abstract}
SPHEREx conducts the first all-sky near-infrared spectral survey over $0.75$--$5.0\um$, spanning the transition from host-galaxy starlight to AGN-heated torus emission and enabling spectroscopic identification of active galactic nuclei (AGNs). We develop a data-driven framework combining weighted non-negative matrix factorization with forward modeling to learn empirical rest-frame templates from 2,777 normal galaxies and 1,064 AGNs in the North Ecliptic Pole (NEP) field. The resulting 14-component dictionary contains seven galaxy templates spanning stellar-population continua and nebular emission and seven AGN templates capturing emission lines and warm and hot torus continua, providing an additive, operational decomposition. Inferred near-infrared host fractions correlate with independent SDSS spectral and HSC imaging decompositions (at Pearson $r=0.68$ and $r=0.59$), comparable to the SDSS--HSC correlation ($r=0.66$) and supporting the reliability of the AGN--host separation. In the NEP deep field, the same templates recover redshifts with $|z_{\rm fit}-z_{\rm spec}|<0.01$ for $93.0\%$ of galaxies and $93.5\%$ of AGNs and distinguish galaxies, narrow-line AGNs, and broad-line AGNs with $81.3\%$ accuracy. At all-sky depth, classification retains $72.9\%$ accuracy. The near-infrared coefficients also flag questionable DESI spectral classifications and redshifts, showing how SPHEREx complements optical spectroscopy. Learned directly from survey spectrophotometry and extending beyond the $K$ band, this empirical dictionary fills a long-standing gap in near-to-mid-infrared templates and provides a data-driven foundation for population-scale studies of AGN activity and its connection to host galaxies across the SPHEREx all-sky survey.
\end{abstract}


\input{01-intro}
\input{02-data}
\input{03-method}
\input{04-results}
\input{05-discussion}
\input{06-conclusion}

\begin{acknowledgments}
We acknowledge support from the National Key R\&D Program of China (No.~2023YFA1607903). 
WKR acknowledges the support from the NSFC grant (Grant No. 12633003).
HXG acknowledges support from the NSFC (Nos.~12473018 and 12522304) and the Overseas Center Platform Projects, CAS (No.~178GJHZ2023184MI).
WWZ is supported by the Shanghai Natural Science Foundation Youth Project (Grant No. 25ZR1402546), Strategic Priority Research Program of the Chinese Academy of Sciences (Grant No. XDB0800302), and the National Key Research and Development Program of China (Grant No. 2025YFA1614102).
MFG is supported by the National Science Foundation of China (grant 12473019), the China Manned Space Project with No. CMS-CSST-2025-A07, the National SKA Program of China (Grant No. 2022SKA0120102), and the Shanghai Pilot Program for Basic Research-Chinese Academy of Science, Shanghai Branch (JCYJ-SHFY-2021-013).
JXW is supported by GSFC (No. 12533006), and Guizhou Provincial Major Scientific and Technological Program XKBF (2025)010 and XKBF (2025)011.

This publication makes use of data products from the Spectro-Photometer for the History of the Universe, Epoch of Reionization and Ices Explorer (SPHEREx), which is a joint project of the Jet Propulsion Laboratory and the California Institute of Technology, and is funded by the National Aeronautics and Space Administration.

This research used data obtained with the Dark Energy Spectroscopic Instrument (DESI). DESI construction and operations is managed by the Lawrence Berkeley National Laboratory. This material is based upon work supported by the U.S. Department of Energy, Office of Science, Office of High-Energy Physics, under Contract No. DE--AC02--05CH11231, and by the National Energy Research Scientific Computing Center, a DOE Office of Science User Facility under the same contract. Additional support for DESI was provided by the U.S. National Science Foundation (NSF), Division of Astronomical Sciences under Contract No. AST-0950945 to the NSF's National Optical-Infrared Astronomy Research Laboratory; the Science and Technology Facilities Council of the United Kingdom; the Gordon and Betty Moore Foundation; the Heising-Simons Foundation; the French Alternative Energies and Atomic Energy Commission (CEA); the National Council of Humanities, Science and Technology of Mexico (CONAHCYT); the Ministry of Science and Innovation of Spain (MICINN), and by the DESI Member Institutions: \url{https://www.desi.lbl.gov/collaborating-institutions}. The DESI collaboration is honored to be permitted to conduct scientific research on I'oligam Du'ag (Kitt Peak), a mountain with particular significance to the Tohono O'odham Nation. Any opinions, findings, and conclusions or recommendations expressed in this material are those of the author(s) and do not necessarily reflect the views of the U.S. National Science Foundation, the U.S. Department of Energy, or any of the listed funding agencies.

The implementation of the analysis code was assisted by Claude Code CLI\footnote{\url{https://code.claude.com/docs/en/overview}}, configured to use a GLM model provided by Z.AI\footnote{\url{https://docs.z.ai/guides/overview/overview}}. OpenAI Codex\footnote{\url{https://openai.com/codex/}} assisted with drafting and language editing of the manuscript. All AI-assisted code and text were reviewed and verified by the authors. All scientific ideas, methodological choices, analyses, interpretations, and conclusions remain the responsibility of the authors.
\end{acknowledgments}

\facilities{SPHEREx, Mayall (DESI)}

\software{Astropy \citep{AstropyCollaboration2013,AstropyCollaboration2018,AstropyCollaboration2022},
NumPy \citep{Harris2020}, SciPy \citep{Virtanen2020}}

\appendix
\renewcommand{\theequation}{\thesection.\arabic{equation}}%
\renewcommand{\theHequation}{\thesection.\arabic{equation}}%
\input{appendix}

\bibliography{refs}{}
\bibliographystyle{aasjournalv7}

\end{document}

%% file: 01-intro.tex
\section{Introduction}
\label{sec:intro}


Understanding the connection between supermassive black hole growth and host-galaxy evolution requires disentangling nuclear accretion emission from stellar emission \citep[e.g.,][]{Kauffmann2003}. Because both the active galactic nucleus (AGN) and the host galaxy contribute to the observed light, this separation is classically approached through multi-wavelength spectral energy distribution (SED) modeling \citep[e.g.,][]{Leja2017, Boquien2019}. Broad-band photometric surveys spanning from the ultraviolet to the infrared have provided the foundation for these efforts by capturing the energy balance between obscured optical/UV light and its thermal re-emission by dust. However, although broad-band SED fitting provides a broad view of this energy balance, its coarse photometric sampling washes out detailed spectral shapes. Consequently, robust decomposition often relies on pre-assumed parametric templates, such as particular stellar population synthesis and torus models. These templates can introduce systematic biases and physical degeneracies if the intrinsic spectral diversity falls outside their assumptions \citep{Leja2018, Yang2020}. Empirical galaxy and AGN SED templates provide an alternative but have likewise been constrained primarily by broad-band photometry \citep[e.g.,][]{Assef2010}.

Spectroscopy resolves these degeneracies by capturing the detailed spectral features that distinguish physical emission mechanisms. Historically, large optical spectroscopic surveys, most notably SDSS \citep{Shen2011, Sexton2021, Ren2024}, have provided the empirical baseline for AGN population studies. Optical spectroscopic AGN samples nevertheless remain incomplete because optical target selection can miss faint or reddened sources, while dust extinction and host-galaxy dilution can weaken recognizable AGN signatures. A complete demographic census therefore requires near-infrared spectroscopic information complementary to these large optical surveys, particularly where hot dust from the AGN torus begins to dominate over the stellar continuum \citep{Hickox2018}. Unfortunately, because ground-based observations are severely hampered by atmospheric thermal backgrounds, systematic, large-scale spectroscopy across the rest-frame near-infrared has largely been out of reach. Targeted JWST observations now probe parts of this wavelength range \citep[e.g.,][]{McKinney2026} but remain limited to selected samples. Population-scale coverage therefore remains a major gap in empirical infrared AGN--host decomposition.

SPHEREx is conducting the first all-sky spectral survey in the near-infrared, covering $0.75{-}5.0\,\mu\mathrm{m}$ with native resolving power $R \sim 35{-}130$ through a system of linear variable filters (LVFs) mounted on six focal-plane arrays \citep{Bock2026, Hui2026}. This observing mode provides uniform sky coverage and large statistical samples while retaining wavelength information across the spectral region. Early-science studies have used broad emission lines in SPHEREx spectra to confirm high-redshift quasars \citep{Davies2026} and heavily reddened quasars \citep{Stepney2026}. This statistical capability opens up a new window to systematically explore the demographics of the missing accreting population.

Beyond source identification, the scientific opportunity is to use these large statistical samples to study how near-infrared continuum shapes and emission-line strengths vary across galaxies and AGNs, and to express those trends in a compact empirical basis. We frame this problem specifically as AGN--host decomposition: we seek a representation that can separate host-galaxy and nuclear contributions in a statistically consistent way while remaining faithful to the native SPHEREx sampling.

This naturally suggests feature-extraction approaches that describe spectral diversity with a compact empirical basis. Principal component analysis (PCA), for example, has been highly successful in optical surveys, where it has been used for source classification \citep{Bolton2012, Guy2023} and host-galaxy decomposition \citep[e.g.,][]{Yip2004, Yip2004a, VandenBerk2006, Shen2011}. By construction, PCA compresses a large data set into a small number of eigenspectra and thus provides an efficient framework for statistical spectral analysis.

However, the SPHEREx data structure differs in an important way from the gridded spectra to which standard PCA is usually applied. Each source is measured as a sparse sequence of low-resolution photometric points in the observed frame, and the source-dependent LVF sampling naturally introduces sub-channel shifts in effective wavelength coverage across the survey. Standard PCA assumes a well-gridded input matrix defined on a common wavelength grid, so applying it directly would require rebinning the data onto a shared set of channels. That step would erase the dither-like wavelength information carried by the ensemble and forfeit the latent spectral resolution available in the survey.

In this work, we construct an empirical rest-frame spectral dictionary directly from SPHEREx spectrophotometry and use it to separate host-galaxy and AGN emission. We cast template recovery as a data-driven dictionary learning problem \citep{Mairal2009,Beckouche2013} and use weighted non-negative matrix factorization (WNMF) to learn both the templates and their additive coefficients under non-negativity constraints and heteroscedastic measurement uncertainties \citep{Lee1999,Blanton2007}. The resulting basis is learned from the observations without pre-computed stellar-population or torus templates.

To preserve the ensemble wavelength information, we forward-model the dictionary through the redshift-dependent response matrix of each source rather than rebinning the measurements onto common channels. Our two-stage strategy first learns a galaxy dictionary from normal galaxies, then holds it fixed while learning additional components from the AGN sample. This focuses the additional basis vectors on spectral variation required by the active population and yields a compact representation for AGN--host decomposition and downstream population studies.

This paper is organized as follows. Section~\ref{sec:data} describes the data and sample construction. Section~\ref{sec:method} presents the WNMF framework and two-stage training. Section~\ref{sec:results} characterizes and validates the recovered components. Section~\ref{sec:applications} demonstrates applications to AGN--host decomposition, redshift estimation, and source classification, and discusses the limitations. Section~\ref{sec:conclusion} summarizes our conclusions. Throughout this work, we adopt a flat $\Lambda$CDM cosmology with $H_0=70\,\mathrm{km\,s^{-1}\,Mpc^{-1}}$, $\Omega_{\rm m}=0.3$, and $\Omega_\Lambda=0.7$.

%% file: 02-data.tex
\section{Data and Sample}
\label{sec:data}

\subsection{The SPHEREx Observatory}
\label{ssec:spherex}

The Spectro-Photometer for the History of the Universe, Epoch of Reionization, and Ices Explorer \citep[SPHEREx, ][]{Bock2026} is a NASA Medium-Class Explorer mission conducting the first all-sky near-infrared spectral survey. It launched on 2025 March 11 and began regular science operations on 2025 May 1.
SPHEREx employs linear variable filters (LVFs), whose passband center varies smoothly along one spatial axis of each detector, to obtain low-resolution spectrophotometry across $0.75$--$5.0\,\mu\mathrm{m}$ in 102 spectral channels at $R \approx 35$--$130$ \citep{Hui2026}.
Because the sampled wavelength depends on detector location, the measurements form an irregular spectral sequence rather than a uniformly sampled spectrum.
The point-spread function has ${\rm FWHM}\sim6''$, with an undersampled pixel scale of $6\farcs15\,\mathrm{pixel}^{-1}$ \citep{Crill2020}. This PSF size is similar to WISE $W1$ and $W2$, whose FWHM values are $6\farcs1$ and $6\farcs4$, respectively \citep{Wright2010}.

The 25-month nominal survey produces four complete all-sky passes with $5\sigma$ point-source depths of $19.5$--$19.9$\,mag (AB) at $\lambda < 3.8\,\mu\mathrm{m}$ and $17.8$--$18.8$\,mag (AB) at $3.8 < \lambda < 5.0\,\mu\mathrm{m}$ \citep{Bock2026}.
The Sun-synchronous orbit naturally overlaps at the ecliptic poles, creating two deep fields of $\sim\!100\,\mathrm{deg}^{2}$ each.
Centered on the North Ecliptic Pole (${\rm RA}=270\fdg0$, ${\rm Dec}=+66\fdg5$), the NEP deep field contains a $1\fdg75$-radius core that receives ${>}\,400$ complete spectral visits over the full 25-month mission \citep{Feder2024}.
We construct the dictionary from the first year of observations within this core, corresponding to ${\sim}\,200$ visits per channel.
The dense repeat coverage provides high per-channel S/N, fills in LVF spectral gaps through dithering, and suppresses detector-level systematics. These properties are essential for recovering weak AGN spectral features in the decomposition (\S\ref{sec:method}; see Figure~\ref{fig:quasar_sed_lc} for an illustrative example).

For the transfer experiments in Section~\ref{sec:applications}, we define an equal-area shallow comparison field of radius $1\fdg75$ centered at ${\rm RA}=215\fdg5$ and ${\rm Dec}=52\fdg5$, near the well-studied Extended Groth Strip \citep{Davis2007}. Hereafter, we refer to this custom footprint as the EGS field. The region is covered by 26 DESI tiles, matching the tile count within the NEP field and thus providing a comparable density of spectroscopic anchors over the same sky area. The field has the nominal SPHEREx all-sky visit coverage and therefore serves as a shallow-field reference for evaluating the transfer of the NEP-trained dictionary.

\subsection{DESI Spectroscopic Anchors}
\label{ssec:desi}

We use spectroscopic data from the Dark Energy Spectroscopic Instrument \citep[DESI, ][]{Levi2019}, a wide-field, multi-object spectrograph that observes through $1\farcs5$-diameter fibers on the 4\,m Mayall telescope at Kitt Peak National Observatory.
DESI positions 5000 robotic fibers across a $3\fdg2$ diameter field of view and covers $3600$--$9800$\,\AA\ at $R \approx 5000$.
The main survey footprint includes substantial coverage of the NEP region; we use DESI Data Release~1 \citep[][hereafter DESI]{AbdulKarim2026}.

DESI provides two ingredients essential for this work: precise spectroscopic redshifts to transform SPHEREx photometry into the rest frame, and comprehensive spectroscopic classifications to separate AGN from normal galaxies.
We adopt the DESI AGN/Galaxy Classification Value-Added Catalog\footnote{\url{https://data.desi.lbl.gov/doc/releases/dr1/vac/agngal/}}, which aggregates a comprehensive suite of optical and mid-infrared AGN diagnostics into a unified bitmask (\texttt{AGN\_MASKBITS}) for each DESI source.
These include, among others, broad emission-line detection (Type-1 AGN), multiple narrow-line ratio methods such as the BPT diagram \citep{Baldwin1981, Kewley2001}, WHAN \citep{CidFernandes2011}, and the mass-excitation diagram \citep{Juneau2014}, as well as WISE color criteria \citep{Stern2012, Jarrett2011, Assef2018}.
The VAC also provides emission-line measurements for all classified sources (S. Juneau et al. 2026, in preparation).
To ensure the purity of the normal-galaxy training sample, we conservatively exclude any source flagged by even a single diagnostic ($\texttt{AGN\_ANY}$ is true), reserving for galaxy template training only sources with no adopted AGN diagnostic flag.

From the DESI catalog we construct a parent sample with secure, unique redshift measurements. We require $\texttt{zcat\_primary} = \mathrm{True}$ to select the catalog-recommended primary coadded spectrum when a target has multiple entries and $\texttt{zwarn} = 0$ to exclude known redshift-fitting failures. We further restrict the spectral class to sources labeled as GALAXY or QSO.
We also require $m_{\mathrm{AB}, W1} < 18.5$ in the WISE~$W1$ band, using the $3.4\,\mu\mathrm{m}$ continuum brightness as a practical proxy for SPHEREx detectability. Because SPHEREx is least sensitive at the longest wavelengths \citep{Bock2026}, this cut reduces the loss of long-wavelength measurements and helps preserve complete $0.75$--$5.0\,\mu\mathrm{m}$ spectral coverage. The resulting parent sample contains $7{,}433$ sources.

\subsubsection{Confusion Screening}
\label{sssec:deblend}

Source confusion is a central challenge for SPHEREx spectrophotometry \citep{Huai2026}.
The instrumental PSF FWHM varies across the spectral channels from approximately $4\farcs6$ at the shortest wavelengths to $6\farcs8$ in the longest-wavelength detector\footnote{\url{https://github.com/SPHEREx/Public-products/blob/master/FWHM_v28_base_cbe.txt}}, and the photometric aperture adopted for spectral sequence construction extends to a diameter of $2.5\times\mathrm{FWHM}$, encompassing a sky area large enough to include unrelated sources.
To ensure that the empirical templates are trained on minimally contaminated photometry, we screen out targets affected by close pairs or extended foreground wings. This quality-control step uses higher-resolution reference catalogs to identify and reject potentially contaminated targets.

We first exploit the superior angular resolution of the Euclid catalog in the NEP region \citep{EuclidCollaboration2025,Jiang2026} to establish a reliable near-infrared counterpart for each DESI target and to identify neighboring sources within the photometric aperture.
Targets are rejected if any companion within the $2.5\times\mathrm{FWHM}$-diameter aperture has $\Delta m_\mathrm{H}<2.5$. This criterion corresponds to a companion-to-target H-band flux ratio greater than 0.1 and reduces the sample to $4{,}054$ sources.

The Euclid catalog is incomplete for bright sources due to saturation, so we turn to the Two Micron All Sky Survey \citep[2MASS;][]{Skrutskie2006} to assess foreground contamination from nearby stars.
For each target, we evaluate the aggregate contamination from 2MASS sources using a double-Gaussian PSF model that captures both the core and the extended wings of the SPHEREx profile.
Targets are discarded when the modeled fractional contamination within the photometric aperture exceeds $5\%$, leaving $3{,}841$ sources: $2{,}777$ normal galaxies and $1{,}064$ with at least one AGN diagnostic flag in the DESI VAC.

\begin{figure}[ht!]
\plotone{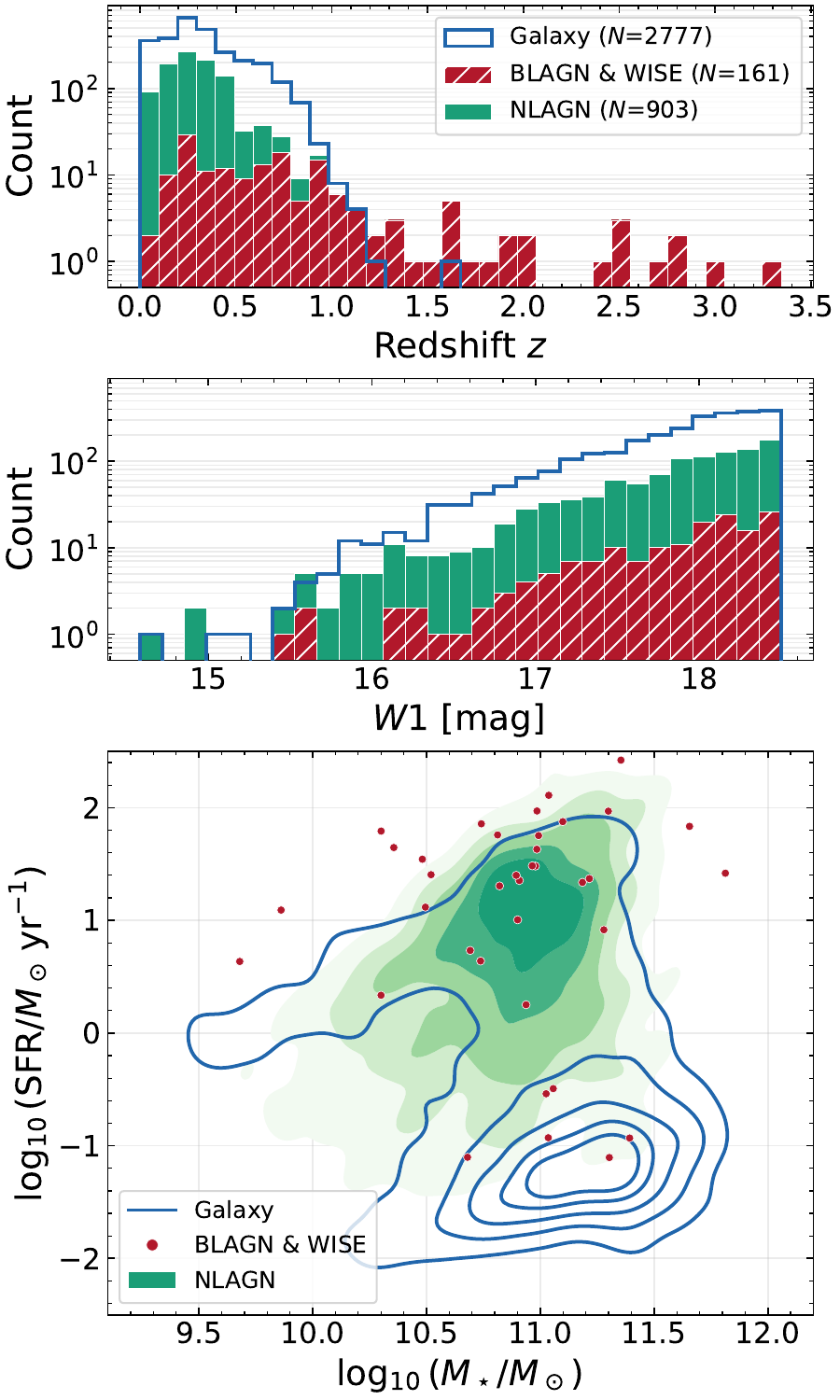}
\caption{%
  Distributions of spectroscopic redshift (top), WISE $W1$-band magnitude (middle), and SFR versus $M_\star$ (bottom) for the selected sample. The sample is broadly separated into normal galaxies (blue step histogram) and AGNs (stacked filled histograms). The AGN population is further subdivided into BLAGN/WISE (sources selected from broad emission lines or WISE colors) and NLAGN/other (sources identified by the remaining diagnostic flags). The bottom panel shows the $M_\star$--SFR plane for sources with valid CIGALE SED measurements, using blue contours for normal galaxies, red points for BLAGN/WISE, and green filled contours for NLAGN/other. Targets without valid SED measurements are omitted from the bottom panel.
}
\label{fig:sample_dist}
\end{figure}

Figure~\ref{fig:sample_dist} illustrates the redshift, WISE $W1$-band magnitude, and $M_\star$--SFR distributions of the selected sample, which we broadly divide into normal galaxies and AGNs.
Within the AGN population, broad-line and WISE-selected sources overlap strongly in their near-infrared component distributions and both show prominent torus emission. We therefore group them as BLAGN/WISE for the population comparisons below, while the remaining AGNs are collectively labeled as NLAGN/other.
The normal-galaxy and NLAGN/other populations are predominantly found at $z \lesssim 1$, whereas the BLAGN/WISE sample extends to much higher redshifts ($z > 3$).
The source counts for all sub-populations rise steadily toward our adopted brightness limit of $W1 = 18.5$\,mag.

The bottom panel compares the $M_\star$--SFR distributions of all three populations using estimates from the DESI DR1 Stellar Mass and Emission Line Value-Added Catalog \citep{Zou2024}, whose SED fits include both galaxy and AGN emission. The normal galaxies are concentrated in the quiescent locus, whereas NLAGN/other sources lie predominantly along the star-forming main sequence and in the green valley. The two distributions nevertheless overlap substantially. In particular, the lower-density normal-galaxy contours extend into the star-forming locus, so the galaxy training sample includes direct examples of the stellar, nebular, and dust-emission variation relevant to AGN hosts. The BLAGN/WISE sources with valid SED measurements span both the star-forming and quiescent regions and overlap the other two populations.
Although the three populations have different density distributions, their substantial overlap in the $M_\star$--SFR plane indicates that the normal-galaxy training sample provides the coverage of host-galaxy properties needed to represent AGN hosts.

\subsection{SPHEREx Photometry and Sequence Construction}
\label{ssec:spherex-phot}

We extract per-channel photometry directly from SPHEREx Level-2 Multi-Extension FITS (MEF) spectral images generated by the official image and spectrophotometry processing pipeline \citep{Akeson2026, Cukierman2026}. The data span observations from 2025 Week~17 through 2026 Week~13.
For each target, we project the DESI spectroscopic position onto the detector pixel grid and extract a $30 \times 30$\,pixel cutout centered on the source, within which we perform all subsequent background subtraction and flux measurements.

The sky background within each cutout consists of a smooth zodiacal emission component \citep{Crill2025} and stochastic residuals from detector noise and potential unresolved faint sources.
We model and subtract the zodiacal component by fitting a scalar scaling factor between the image and the pipeline zodiacal template via inverse-variance-weighted least squares.
This fit uses background pixels selected after excluding those within the photometric aperture and those flagged for data-quality issues (e.g., saturation or cosmic rays).
The residual background level and its uncertainty are then estimated as the sigma-clipped median and standard error of these selected pixels.

Source photometry is extracted within circular apertures whose diameter is set to $2.5 \times \mathrm{FWHM}$ for each spectral image. The aperture therefore varies with the channel-dependent PSF, with radii of approximately $0.9$--$1.4$\,pixels across the $4\farcs6$--$6\farcs8$ FWHM range at $6\farcs2\,\mathrm{pixel}^{-1}$.
The background-subtracted fluxes are converted to $\mu$Jy, with photometric uncertainties formally propagated from the detector variance and residual background error.
For each measurement, the \texttt{FLAGS} bits from all pixels within the aperture are combined through a bitwise OR, so the resulting flag records any detector-quality condition present in the aperture. Measurements containing designated bad bits are removed before template training.
By incorporating the effective wavelength and bandwidth derived from the local LVF response \citep{Hui2026}, we construct a complete per-visit photometric record (flux, error, flag, $\lambda_{\mathrm{cen}}$, $\Delta\lambda$) for each channel.
The entire data reduction and extraction pipeline is implemented in the open-source \texttt{spxquery} framework\footnote{\url{https://github.com/WenkeRen/spxquery}}.

\begin{figure*}[ht!]
\plotone{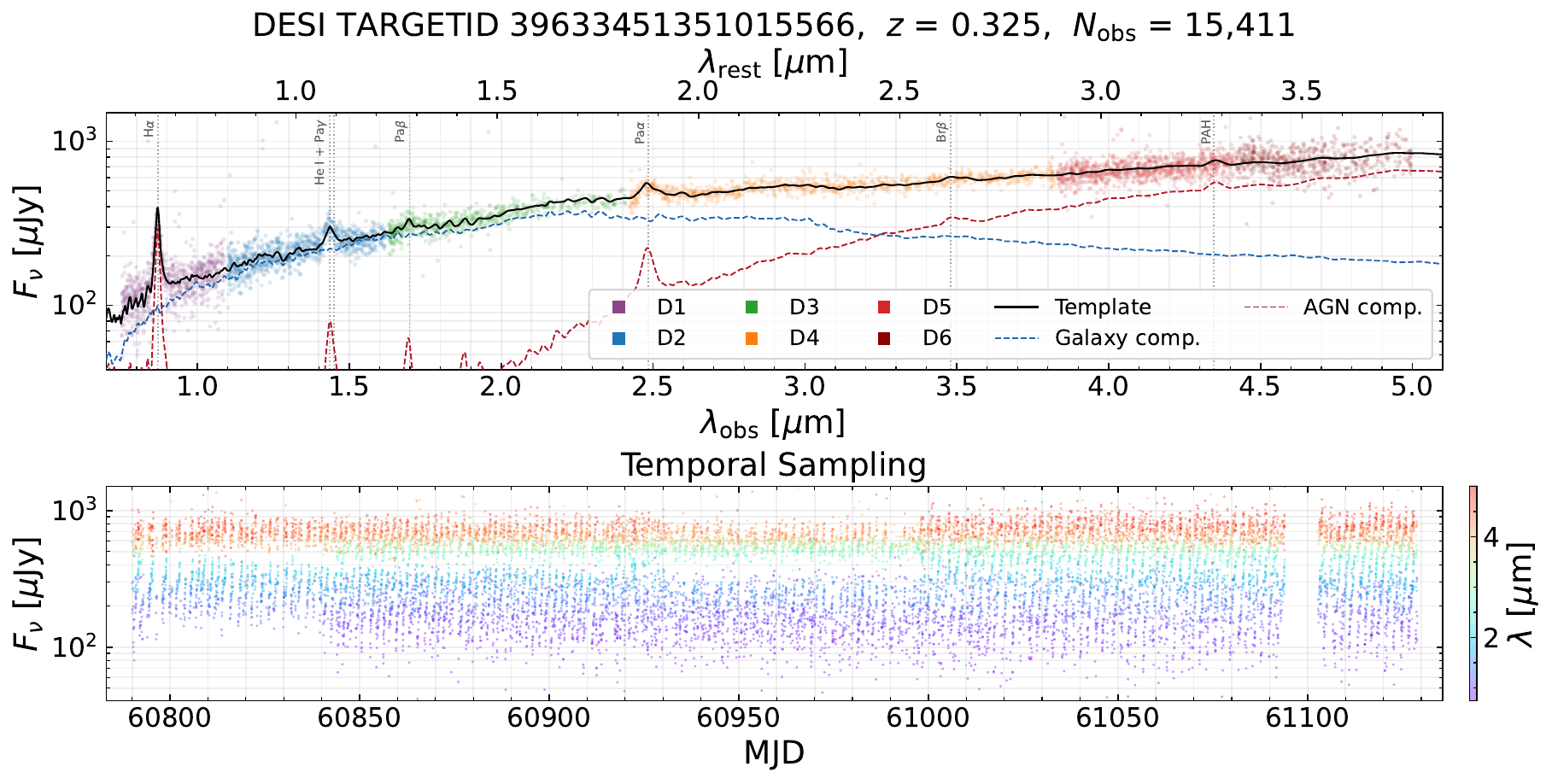}
\caption{%
  SPHEREx spectrophotometry of a representative DESI-identified quasar
  ($z = 0.325$).
  \textit{Top:} observed-frame SED compiled from individual flux
  measurements, color-coded by SPHEREx detector band (D1--D6).
  Solid and dashed curves show the best-fit reconstruction and its decomposed AGN and host-galaxy components (see \S\ref{sec:method}).
  Vertical dotted lines mark the identifiable emission lines; the top axis displays rest-frame wavelengths.
  \textit{Bottom:} individual flux measurements as a function of observing date,
  color-coded by observed wavelength to illustrate the wavelength-dependent
  temporal sampling of the NEP deep field over ${\sim}\,300$\,days.
}
\label{fig:quasar_sed_lc}
\end{figure*}

Figure~\ref{fig:quasar_sed_lc} illustrates the resulting spectrophotometric data products for a representative DESI-identified quasar at $z = 0.325$.

%% file: 03-method.tex
\section{Methodology}
\label{sec:method}

In this section, we describe the data-driven framework used to extract representative spectral features and to decompose the composite spectra of active galaxies into host-galaxy and AGN components. Given the clean, redshift-identified sample defined in Section~\ref{sec:data}, the central methodological challenge is to recover non-parametric spectral structure from sparsely sampled SPHEREx data while retaining a constrained, additive representation.

We address this problem with WNMF \citep{Lee1999,Blanton2007}, implemented here in a two-stage framework. Under the assumption of linear additivity, WNMF learns non-negative spectral templates directly from the data, avoiding restrictive parametric prescriptions for either the stellar or AGN component. The method uses the statistical power of the full sample to capture the dominant modes of spectral variation while retaining an additive representation that can be compared with spectral features and independently measured source properties.

\begin{figure*}[ht!]
    \includegraphics[width=\textwidth]{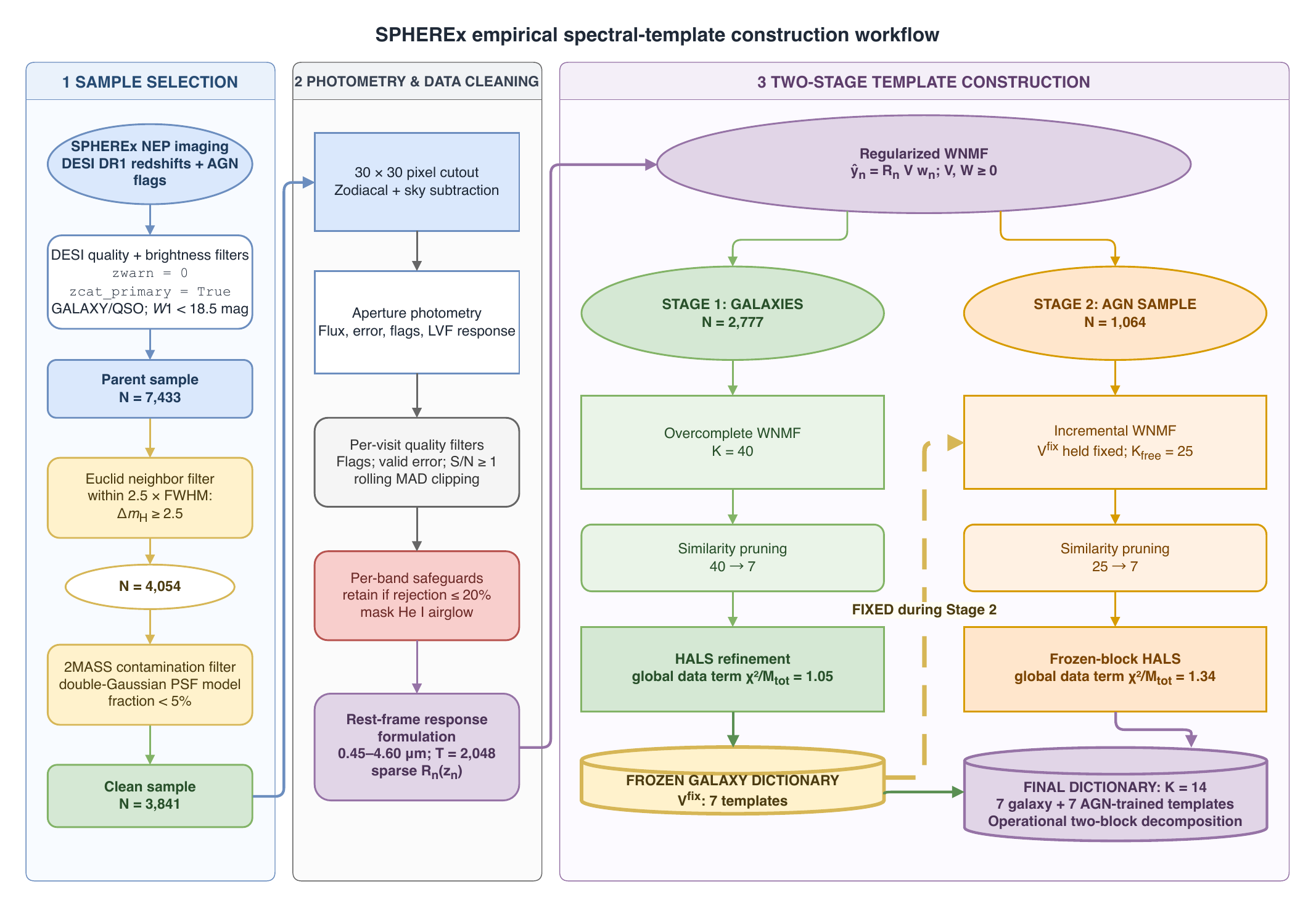}
    \caption{Overview of the empirical spectral-template construction workflow. After sample selection, the SPHEREx photometry is cleaned and mapped to a common rest-frame grid through $\mathbf{R}_n(z_n)$. Seven galaxy templates are learned and frozen as $\mathbf{V}^{\mathrm{fix}}$. In the second stage, these components remain in the fit to account for the stellar host-galaxy contribution, while seven additional templates are learned from the residual spectral variation in the AGN sample, yielding the final $K=14$ dictionary.}
    \label{fig:template_workflow}
\end{figure*}

The sequential design of the framework is central to the decomposition. Figure~\ref{fig:template_workflow} summarizes the path from sample selection, photometric data cleaning, and rest-frame response formulation to the two template-training phases. We first learn a compact galaxy dictionary from the normal-galaxy sample defined in Section~\ref{sec:data}. We then hold this dictionary fixed while learning additional components from the AGN sample. This construction assumes that the host spectra of active systems can be represented by non-negative combinations of the normal-galaxy dictionary, so that the additional components primarily capture residual spectral variation required by the AGN sample. It therefore defines an operational separation between host-associated and AGN-trained spectral variation without imposing parametric spectral templates.

\subsection{Data Preprocessing and Rest-frame Response Formulation}
\label{ssec:restframe}

Starting with the clean, uncontaminated photometric sample constructed in Section~\ref{sec:data}, we apply an additional per-source photometric quality pipeline. We first reject observations with quality bitmasks, invalid uncertainties, or low signal-to-noise ratios ($\mathrm{SNR} < 1$). We perform a rolling Median Absolute Deviation (MAD) sigma-clipping pass to flag transient outliers. To prevent selection bias near the detection limit, where these filters can preferentially reject downward noise fluctuations while retaining upward ones and thus introduce an artificial positive flux bias, we enforce a per-band rejection threshold. If more than $20\%$ of the observations in any given detector band are removed by the preceding steps, we discard all data from that entire band for the source. Finally, we mask observer-frame channels affected by the $\mathrm{He\,I}\ 1.083\,\mu\mathrm{m}$ airglow line.

We define the common rest-frame wavelength domain empirically from the source-coverage distribution after these quality cuts. Of the 3,841 retained sources, requiring each rest-frame wavelength bin to be sampled by at least $5\%$ of the sample (193 sources) yields a contiguous interval of $0.503$--$4.60\,\mu\mathrm{m}$. We extend only the blue boundary to $0.45\,\mu\mathrm{m}$ to retain localized ${\rm H}\beta$ structure at $0.4861\,\mu\mathrm{m}$, despite coverage by only 146 sources ($3.80\%$). The few sources reaching this interval are dominated by higher-redshift AGN. Their sparse coverage can retain the aligned emission-line signal but does not adequately constrain the galaxy continuum below $0.503\,\mu\mathrm{m}$, where artificial continuum fluctuations can arise near the wavelength boundary.

On this basis, we sample the latent spectrum on a common logarithmic grid of $T = 2048$ bins spanning $0.45$--$4.60\,\mu\mathrm{m}$. The constant logarithmic spacing corresponds to a nominal grid resolving power of $R = \lambda / \Delta\lambda \approx 880$, or a velocity sampling of $\approx 340\ \mathrm{km\ s^{-1}}$. This grid is finer than the native SPHEREx channels so that measurements obtained at slightly different LVF samplings across sources and visits can be combined without first rebinning them. It therefore retains the additional spectral information provided by the dithering-like sampling. The nominal $R$ describes the numerical grid rather than the resolution of an individual SPHEREx measurement. Once the response is sampled sufficiently finely, the exact number of grid points has little effect on the recovered templates. For each source $n$ at redshift $z_n$, let $M_n$ denote the number of retained observer-frame flux measurements after the quality cuts above. We define a sparse response matrix $\mathbf{R}_n(z_n) \in \mathbb{R}^{M_n \times T}$ such that a latent rest-frame spectrum $\mathbf{x} \in \mathbb{R}^T$ predicts the measured observer-frame fluxes as $\widehat{\mathbf{y}}_n = \mathbf{R}_n(z_n)\,\mathbf{x}$. The matrix incorporates the redshift and the LVF transmission curves when mapping the latent spectrum to the observed channels. We approximate each passband as a top-hat profile whose central wavelength and bandwidth are determined directly from the Level-2 WCS at the source position.

\subsection{Data-Driven Spectral Decomposition Framework}
\label{ssec:spectral_decomposition}

We formulate our framework as a non-negative dictionary learning problem, with the aim of deriving a compact empirical representation of the spectral diversity resolved by the sparse and irregularly sampled SPHEREx spectrophotometry.

Following this formalism, we represent each source as a non-negative combination of empirical templates. For a sample of $N$ sources, let $\mathbf{y}_n \in \mathbb{R}^{M_n}$ represent the observed flux vector of source $n$, and let $\mathbf{\Sigma}_n = \operatorname{diag}(\sigma_{n,1}^2,\ldots,\sigma_{n,M_n}^2) \in \mathbb{R}^{M_n \times M_n}$ be its noise covariance matrix, where $\sigma_{n,i}$ is the uncertainty of measurement $i$. Under the assumption of linear additivity, we model the latent rest-frame spectrum of the source as a linear combination of $K$ non-negative templates. We define $\mathbf{V} = [\mathbf{v}_1,\ldots,\mathbf{v}_K] \in \mathbb{R}^{T \times K}$ as the rest-frame template dictionary, where $\mathbf{v}_k \in \mathbb{R}^T$ is its $k$-th basis template. The corresponding activation weights are given by the coefficient matrix $\mathbf{W} \in \mathbb{R}^{N \times K}$, where the row $\mathbf{w}_n \in \mathbb{R}^K$ contains the contribution weights for source $n$.

To compare the rest-frame templates directly with the observer-frame measurements, we project them through the response matrix $\mathbf{R}_n(z_n)$ defined in Section~\ref{ssec:restframe}:
\begin{equation}
    \widehat{\mathbf{y}}_n = \mathbf{R}_n(z_n) \, \mathbf{V} \, \mathbf{w}_n.
\end{equation}
The reconstruction error is minimized under non-negativity constraints ($\mathbf{V} \ge 0, \mathbf{W} \ge 0$), yielding an additive representation and preventing cancellations between positive and negative components.

To address the ill-posed nature of this reconstruction from sparse and noisy data, we define the global WNMF objective function:
\begin{equation}
\begin{split}
    \min_{\mathbf{V} \ge 0, \mathbf{W} \ge 0} \mathcal{L} = & \sum_{n=1}^{N} \left( \mathbf{y}_n - \mathbf{R}_n \mathbf{V} \mathbf{w}_n \right)^{\mathsf{T}} \mathbf{\Sigma}_n^{-1} \left( \mathbf{y}_n - \mathbf{R}_n \mathbf{V} \mathbf{w}_n \right) \\
    & + \mathcal{R}(\mathbf{V}),
\end{split}
\end{equation}
where the summed quadratic residual is the data term, denoted $\chi^2$ below, and $\mathcal{R}(\mathbf{V})$ is the regularization term that suppresses unphysical, high-frequency oscillations and stabilizes the template profiles:
\begin{equation}
    \begin{split}
        \mathcal{R}(\mathbf{V}) ={} & \alpha \sum_{k=1}^K \mathcal{R}_{L2}(\mathbf{v}_k)
        + \beta \sum_{k=1}^K \mathcal{R}_{\mathrm{diff}}(\mathbf{v}_k) \\
        & + \gamma \sum_{k=1}^K \mathcal{R}_{\mathrm{curv}}(\mathbf{v}_k).
    \end{split}
\end{equation}
Here, the non-negative hyperparameters $\alpha$, $\beta$, and $\gamma$ control the strengths of the template-amplitude ($L_2$), first-difference (smoothness), and second-difference (curvature) penalties, respectively. Their explicit discretized forms are given in Appendix~\ref{sec:app_optimization}.

We fit the two dictionary blocks sequentially, first on the normal-galaxy sample and then on the AGN sample. Both blocks follow the same three-step architecture: an overcomplete alternating least-squares WNMF (ALS-WNMF) fit explores the spectral variation present in the sample, similarity-based pruning consolidates redundant components, and hierarchical alternating least squares (HALS) refines the resulting compact dictionary. The distinction is that galaxy training updates the entire dictionary, whereas AGN training keeps the learned galaxy block fixed and updates only the additional AGN-trained block.

Galaxy training uses the clean normal-galaxy subset ($N_{\mathrm{gal}} = 2777$) defined in Section~\ref{sec:data}.

We initialize an overcomplete dictionary with $K_{\mathrm{initial}} = 40$ templates, allowing the exploratory fit to represent both common continuum shapes and lower-amplitude spectral structure before the dictionary size is fixed. We minimize the global loss by alternating between non-negative least-squares (NNLS) updates of the source coefficients and multiplicative updates of the templates (Appendix~\ref{sec:app_optimization}). This exploratory fit uses $\alpha = 10^{-3}$ and $\beta = \gamma = 0$, so that only weak $L_2$ regularization is applied.

Second, because over-parameterized training yields redundant templates, we compress the dictionary by merging templates with similar spectral profiles (detailed in Appendix~\ref{sec:app_prune}). The rank-selection tests balance predictive performance on independent visits against component redundancy and coefficient stability. They support retaining seven galaxy templates as a compact representation of the spectral diversity resolved in this sample at the signal-to-noise ratio and spectral resolution of the SPHEREx data (see Appendix~\ref{sec:app_dictionary_size} for details).

Finally, we refine the seven pruned templates with HALS (Appendix~\ref{sec:app_hals}). By updating one template at a time against the residuals left by the others, HALS reduces the tendency of simultaneous updates to produce nearly duplicated profiles. We set $\alpha = 10^{-3}$, $\beta = 1$, and $\gamma = 5$ during this step, using first- and second-difference penalties to prevent noise-driven, high-frequency oscillations from being absorbed into the templates. The resulting galaxy model has a global data term of $\chi^2/M_{\mathrm{tot}}=1.05$, where $M_{\mathrm{tot}}=\sum_n M_n$ is the total number of retained flux measurements in the training sample under consideration. This value is close to unity, indicating that the residuals are commensurate with the reported measurement uncertainties and providing no evidence for substantial global under- or over-fitting. These seven templates form the frozen galaxy dictionary $\mathbf{V}^{\mathrm{fix}}$ used in the subsequent decomposition.

AGN training uses the AGN sample ($N_{\mathrm{AGN}} = 1064$) to learn the additional spectral variation required by active galaxies. We define the joint dictionary as
\begin{equation}
    \mathbf{V} = \left[ \mathbf{V}^{\mathrm{fix}} \;\middle|\; \mathbf{V}^{\mathrm{free}} \right] \in \mathbb{R}^{T \times (K_{\mathrm{gal}} + K_{\mathrm{free}})},
\end{equation}
where $\mathbf{V}^{\mathrm{fix}} \in \mathbb{R}^{T \times K_{\mathrm{gal}}}$ contains the seven frozen galaxy templates, and $\mathbf{V}^{\mathrm{free}} \in \mathbb{R}^{T \times K_{\mathrm{free}}}$ contains the additional templates learned from the AGN sample. The exploratory fit begins with $K_{\mathrm{free}}=25$ (decreased given the smaller sample size) and uses the same weak-$L_2$ setting as the galaxy exploration. The coefficients of both blocks remain free for every source, but only $\mathbf{V}^{\mathrm{free}}$ is updated. In this way, spectral structure already represented by normal galaxies can be assigned to the fixed block, while recurring structure not captured by that block drives the free templates.

We likewise compress the overcomplete free block by merging templates with similar spectral profiles, without pruning or altering the frozen galaxy block. Applying the same rank-selection criteria to the free block supports seven AGN templates (Appendix~\ref{sec:app_dictionary_size}). The equal sizes of the galaxy and AGN blocks are therefore an outcome of the two tests rather than an imposed constraint. We then refine the compact free block with HALS while continuing to hold the galaxy block fixed. This refinement with $(\alpha,\beta,\gamma)=(10^{-3},1,5)$ yields a global data term of $\chi^2/M_{\mathrm{tot}}=1.34$ for the AGN sample. Although higher than the value for normal galaxies, this modest excess is qualitatively consistent with the greater spectral dimensionality of active galaxies found in SDSS PCA analyses: the first three galaxy eigenspectra account for $\approx 98\%$ of the sample variance \citep{Yip2004a}, whereas the first twenty quasar eigenspectra capture $96.9\%$ \citep{Yip2004}. At the spectral resolution of SPHEREx, our independent-visit tests nevertheless show that seven free templates capture the dominant recurring structure and that enlarging the dictionary provides little additional predictive power.

The final $K=14$ dictionary thus contains seven galaxy templates learned from normal galaxies and seven additional templates learned from the AGN sample. Once training is complete, all 14 templates are held fixed and only their coefficients are fitted for each source. The division into galaxy and AGN blocks therefore records their training origin and defines the operational host--AGN decomposition used in this work.

%% file: 04-results.tex
\section{Results}
\label{sec:results}

The two-stage procedure yields a final dictionary of 14 non-negative templates: seven galaxy components learned from the normal-galaxy sample and seven AGN components learned from the AGN sample. We describe the recurring spectral structure represented by this dictionary and test its interpretation against independent measurements.

\subsection{The Empirical Spectral Dictionary}
\label{ssec:empirical_dictionary}

\begin{figure*}[ht!]
    \centering
    \includegraphics[width=\textwidth]{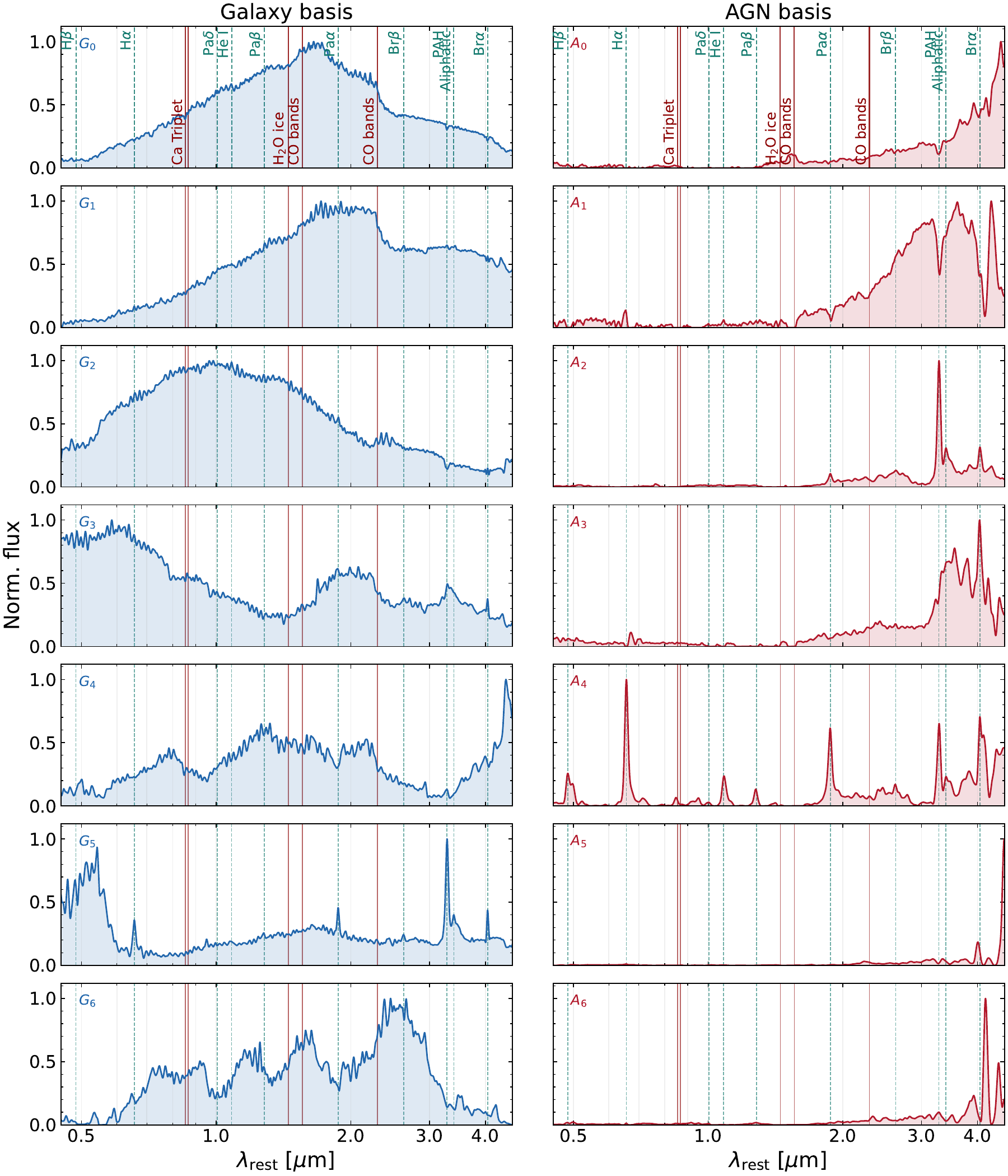}
    \caption{The final $K=14$ empirical spectral dictionary. The left and right columns show the seven galaxy components learned from normal galaxies ($G_0$--$G_6$) and the seven components trained on the AGN sample ($A_0$--$A_6$), respectively. Within each column, the components are ordered from the leading component downward. Each template is peak-normalized for display and shown on a logarithmic wavelength axis. In each panel, emission features are marked by dashed lines, while absorption, break, and ice features are marked by solid lines. Both sets of features are labeled only in the top panel of each column.}
    \label{fig:template_overview}
\end{figure*}

Figure~\ref{fig:template_overview} shows the complete dictionary on the common $0.45$--$4.60\um$ rest-frame grid. The galaxy components contain both smooth stellar-continuum variation and localized structure associated with nebular and dust emission. The AGN components span red continua, shorter-wavelength continuum variation, and components dominated by recombination or PAH features. Because WNMF does not uniquely identify physical source spectra, an individual component need not correspond to a single stellar population, dust temperature, or line-emitting region.

\begin{figure*}[ht!]
    \plotone{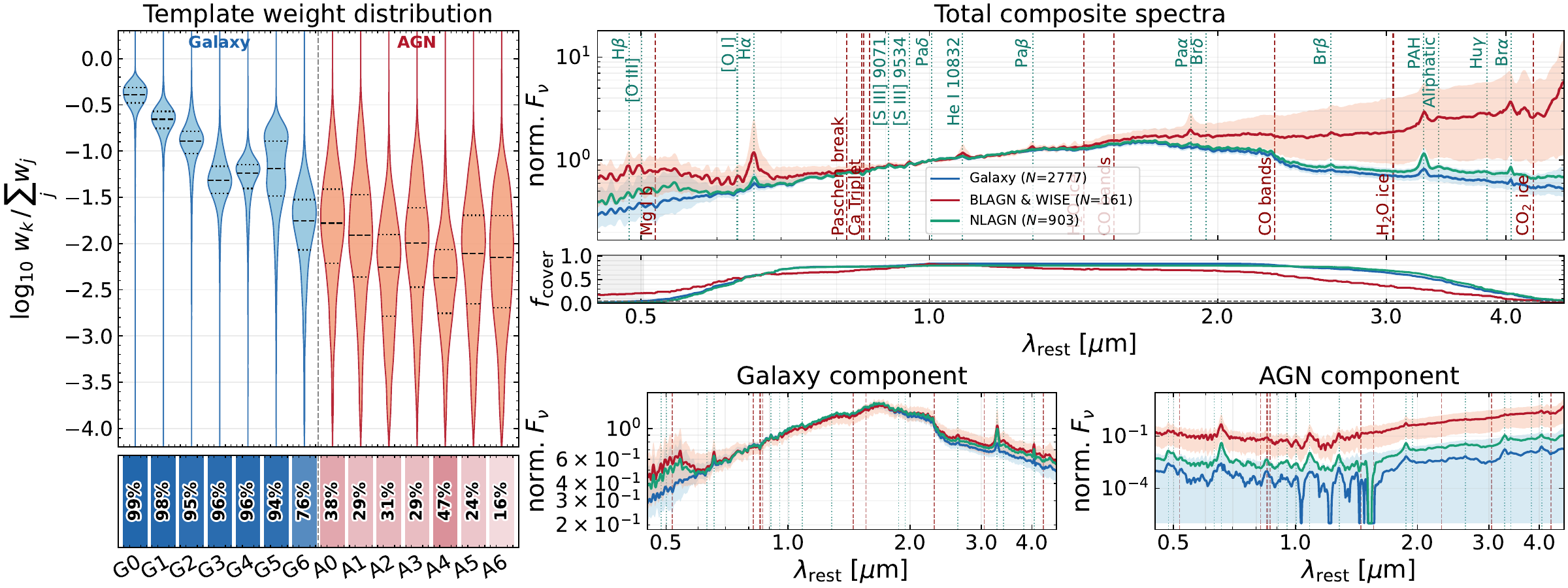}
    \caption{Component weights and reconstructed population composites. \textit{Left:} Distributions of the per-source $L_1$-normalized weights. The strip below gives the fraction of all 3,841 sources for which each normalized weight exceeds $10^{-4}$. \textit{Right:} Median reconstructed spectra for normal galaxies ($N=2{,}777$), the combined broad-line and WISE-selected AGN population ($N=161$), and the remaining NLAGN/other-AGN population ($N=903$). The total spectra are shown above their galaxy- and AGN-component contributions. All three panels use the normalization of the total spectrum at $1\um$ and retain the relative amplitude of the two component groups. The AGN contribution is shown logarithmically. The narrow strip gives the fraction of each population with effective response coverage at both the displayed wavelength and the normalization wavelength. Note that the apparent blue-end rise in the galaxy contribution is an artificial feature caused by the limited coverage explained in Section~\ref{ssec:restframe}.}
    \label{fig:sample_composite}
\end{figure*}

The left side of Figure~\ref{fig:sample_composite} shows the component-weight distributions. Components $G_0$--$G_5$ are active in at least $94\%$ of the full sample, while $G_6$ is active in $76\%$. Together with the overlapping $M_\star$--SFR distributions shown in Figure~\ref{fig:sample_dist}, these activation rates support the interpretation that components learned exclusively from normal galaxies provide a broadly shared host-galaxy representation across the full sample. The AGN components have lower and more heterogeneous activation rates of $16\%$--$47\%$, consistent with spectral structure that is not required universally. Their nonzero activation is not, by itself, an AGN classification criterion because these components can also absorb population differences or features underrepresented in the normal-galaxy training set. We discuss the AGN classification implications of the component weights in Section~\ref{ssec:classification}.

The right side of Figure~\ref{fig:sample_composite} compares the reconstructed population composites. The normal-galaxy spectrum is almost entirely described by the galaxy components. The BLAGN/WISE composite has the strongest short-wavelength continuum and the most prominent rise beyond approximately $2\um$. The NLAGN/other-AGN composite is more host dominated, but its weaker AGN contribution still rises toward longer wavelengths, indicating a hot-dust contribution and a continuum shape distinct from that of normal galaxies. Recombination and PAH structure are also visible in the AGN-component composites. Together, these trends reveal a clear division of roles in the two-stage model: the galaxy components describe spectral variation shared with normal galaxies, whereas the AGN components add structure characteristic of the AGN sample. This additional structure can arise from both nuclear emission and circumnuclear or host-scale dust and line emission associated with active-galaxy environments.

\subsection{Empirical Validation of the Galaxy Components}
\label{ssec:galaxy_validation}

\begin{figure*}[ht!]
    \plotone{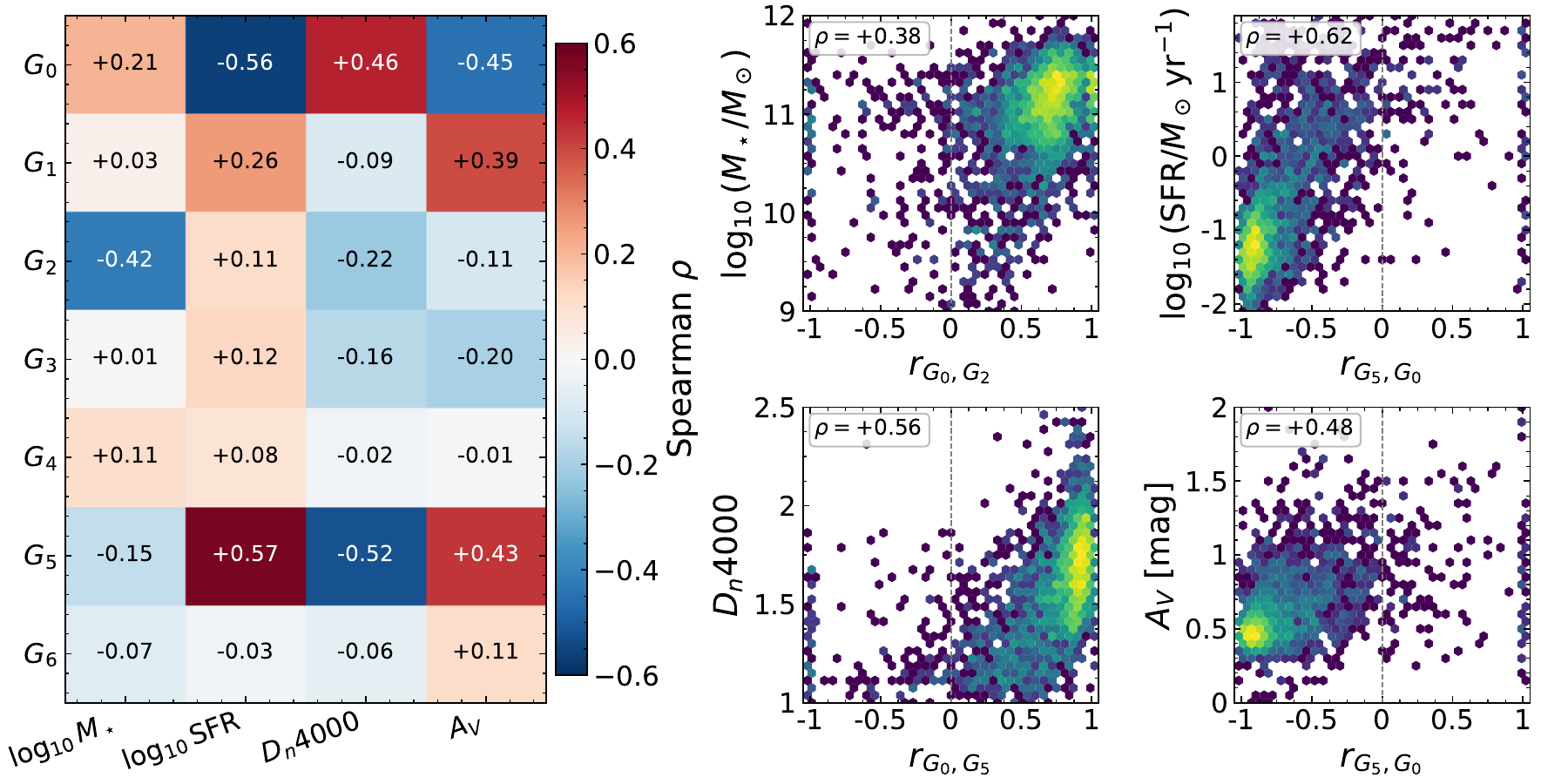}
    \caption{Relation between the galaxy-component coefficients and independently measured galaxy properties. \textit{Left:} Spearman rank correlations between the normalized galaxy-component weights and stellar mass, SFR, $D_n4000$, and $A_V$. \textit{Right:} Normalized contrasts $r_{i,j}=(w_i-w_j)/(w_i+w_j)$ between the most positively and negatively correlated components, where $w_i$ and $w_j$ are their normalized weights. For every contrast, $r_{i,j}=+1$ denotes exclusive contribution from the first named component and $r_{i,j}=-1$ exclusive contribution from the second. We show $r_{G0,G2}$ for stellar mass, $r_{G5,G0}$ for SFR, $r_{G0,G5}$ for $D_n4000$, and $r_{G5,G0}$ for $A_V$; Spearman coefficients and sample sizes are given in each panel. Stellar mass, SFR, and $A_V$ are obtained from DESI Value-Added Catalog SED fits, whereas $D_n4000$ is measured from the optical spectrum \citep{Zou2024}.}
    \label{fig:property_correlation}
\end{figure*}

To investigate the physical variation encoded by the individual galaxy components, we compare their weights for the 2,777 normal galaxies with stellar mass, SFR, dust attenuation ($A_V$), and $D_n4000$ from the DESI DR1 Stellar Mass and Emission Line Value-Added Catalog \citep{Zou2024}. The first three quantities are obtained from CIGALE SED fitting, whereas $D_n4000$ is measured from the rest-frame optical spectrum. The narrow 4000~\AA\ break increases with the luminosity-weighted age of a stellar population, although it also depends partly on metallicity and star-formation history \citep{Balogh1999}.

The left side of Figure~\ref{fig:property_correlation} shows that $G_0$ has the broadest pattern of association with the independently inferred physical properties. Its weight increases with $D_n4000$ and stellar mass but decreases with SFR and $A_V$, identifying it as the backbone of an older, less actively star-forming stellar continuum. In contrast, $G_1$ increases with both $A_V$ and SFR, tracing a dusty star-forming continuum with little dependence on stellar mass or $D_n4000$. Most notably, $G_5$ has the strongest positive association with SFR and a comparably strong decrease with $D_n4000$, while also increasing with $A_V$. Its numerous PAH and recombination features are therefore consistent with dust-rich star-forming and nebular emission. Component $G_2$ shows the strongest anticorrelation with stellar mass ($\rho=-0.42$). Together with its comparatively blue continuum (Figure~\ref{fig:template_overview}), this trend is consistent with a larger fractional contribution from young, low-mass-to-light-ratio stellar populations in lower-mass galaxies. The correlations of $G_3$, $G_4$, and $G_6$ are comparatively modest.

To show how relative component contributions track these population changes, the right side of Figure~\ref{fig:property_correlation} presents the normalized weight contrast $r_{i,j}=(w_i-w_j)/(w_i+w_j)$, where $w_i$ and $w_j$ are the normalized weights of components $i$ and $j$. For each property, we pair the components with the most positive and negative correlations in the left panel. The $G_0$--$G_5$ balance traces SFR ($\rho=0.62$), $D_n4000$ ($\rho=0.56$), and $A_V$ ($\rho=0.48$), linking the old-stellar backbone to the dusty star-forming and nebular contribution. Stellar mass is instead traced by the $G_0$--$G_2$ contrast ($\rho=0.38$), consistent with a shift from the old-stellar backbone toward a larger fractional contribution from the blue, young stellar continuum of $G_2$. Because these physical quantities are inferred independently of the SPHEREx decomposition, and $D_n4000$ is measured outside the template wavelength range, the correlations provide an independent validation that the learned components encode astrophysically meaningful population variation. They further indicate that fitted weight contrasts can serve as empirical proxies for these galaxy properties.

\begin{figure}[ht!]
    \plotone{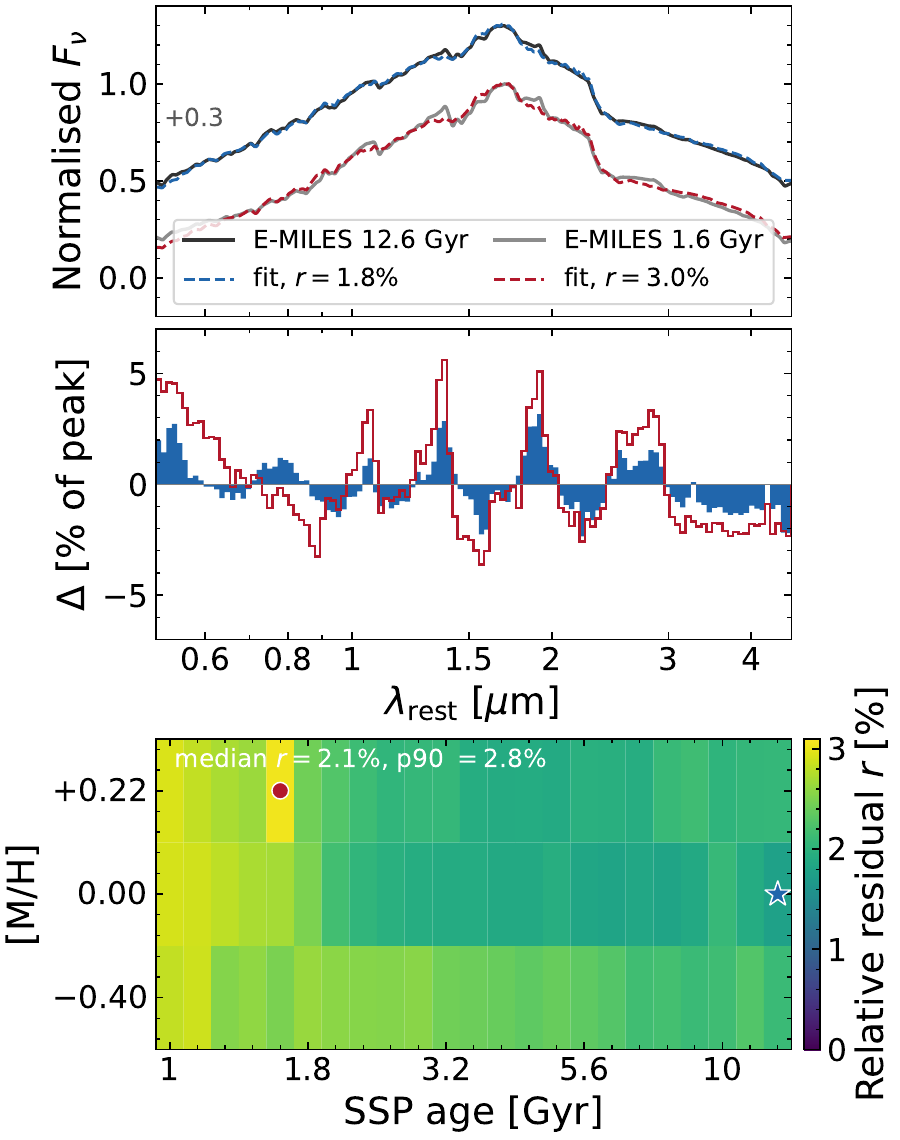}
    \caption{Projection of 69 unattenuated E-MILES SSPs onto the seven galaxy components over $0.505$--$4.60\um$. \textit{Top:} Best-fit (age $12.6$~Gyr, $[\mathrm{M/H}]=0.00$, $r=1.8\%$) and worst-fit (age $1.6$~Gyr, $[\mathrm{M/H}]=+0.22$, $r=3.0\%$) SSPs and their reconstructions; the former is offset by $0.3$. \textit{Middle:} Data-minus-model residuals in percent of the target peak. \textit{Bottom:} Residual $r$ across the age--metallicity grid, with the two examples marked and the sample median and 90th percentile annotated.}
    \label{fig:ssp_projection}
\end{figure}

The component--property correlations establish that the galaxy dictionary follows population changes within the observed sample. We next test whether the same seven-component representation also spans standard stellar continua over a controlled age--metallicity grid. We fit non-negative combinations of the galaxy components to 69 unattenuated E-MILES SSP spectra \citep{Rock2016, Vazdekis2016} with ages of $1.0$--$13.5$~Gyr and $[\mathrm{M/H}]=-0.40$, $0.00$, and $+0.22$ (Figure~\ref{fig:ssp_projection}). The fits cover $0.505$--$4.60\um$, where the source coverage provides reliable constraints on the galaxy continuum. As described in Section~\ref{ssec:restframe}, the template grid extends farther blueward to $0.45\um$ to retain localized ${\rm H}\beta$ structure, but the sparse sampling below approximately $0.503\um$ does not constrain the galaxy continuum reliably. We therefore exclude that blue extension from this SSP coverage test.

The reconstructions follow the continuum structure across the full grid, with a median coverage-weighted relative residual of $2.1\%$ and a 90th percentile of $2.8\%$. Even the largest residual is $3.0\%$ and occurs for a young, super-solar-metallicity SSP. The low residuals demonstrate that the galaxy components provide an accurate and compact representation of standard stellar-population continua across this parameter range. The empirical dictionary additionally retains observed nebular and PAH contributions absent from E-MILES, most clearly through $G_5$.

\subsection{Diagnostics of the AGN Components}
\label{ssec:agn_validation}

\begin{figure*}[ht!]
    \plotone{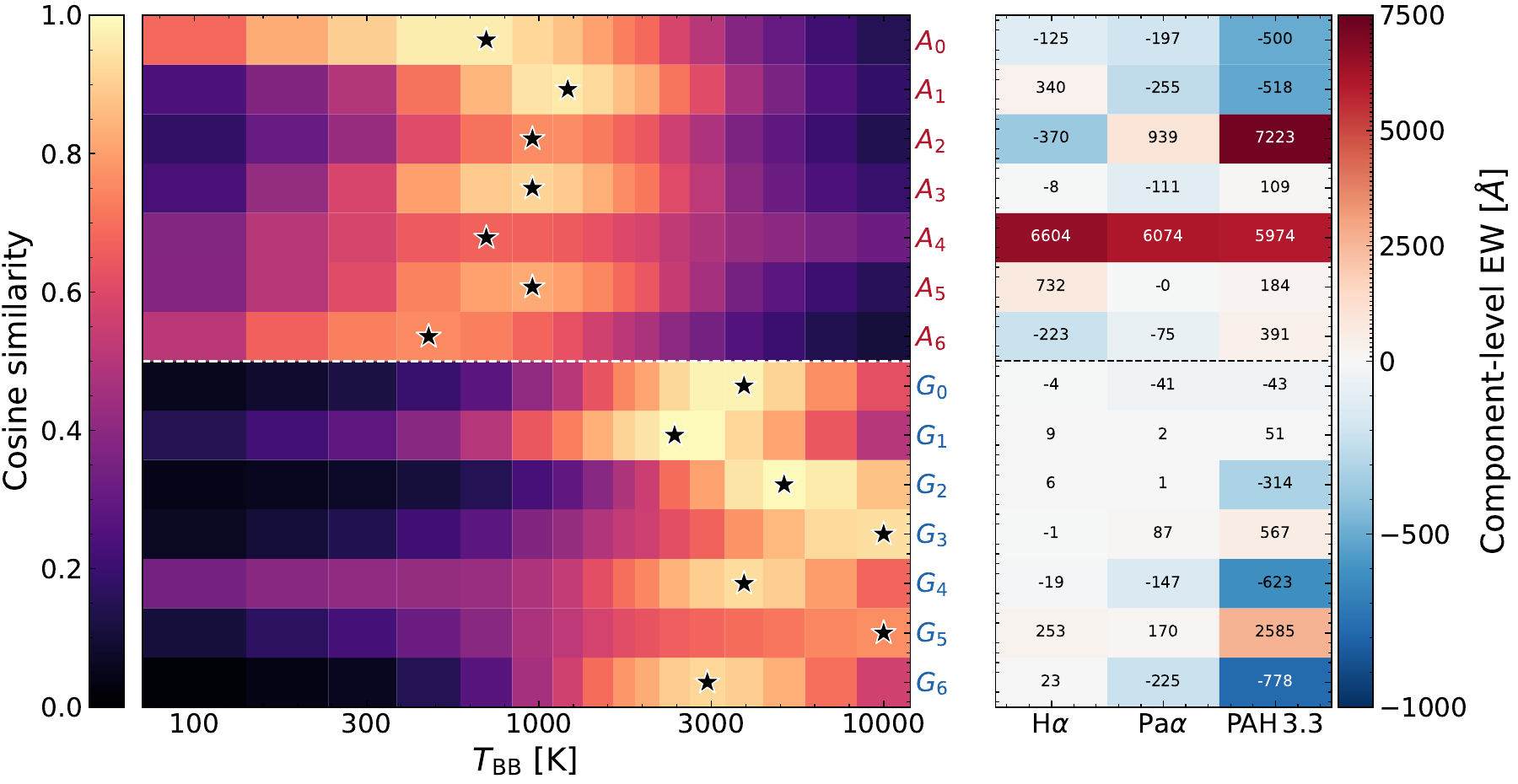}
    \caption{Continuum-shape and emission-feature diagnostics for the AGN (top) and galaxy (bottom) components. \textit{Left:} Cosine similarity to blackbody spectra over $T_{\rm BB}=100$--$10{,}000\,\mathrm{K}$, using the metric defined in Appendix~\ref{sec:app_prune}. Stars indicate the maximum similarity in each row. The maximizing values provide a common color-temperature coordinate for comparing continuum shapes. The AGN components peak near $1000\,\mathrm{K}$, characteristic of hot dust, whereas the galaxy components peak at $3000$--$10{,}000\,\mathrm{K}$, characteristic of stellar continua. \textit{Right:} Component-level EWs of ${\rm H}\alpha$, ${\rm Pa}\alpha$, and the PAH $3.3\um$ feature after broadening to $R=50$ and fitting a local linear continuum. Positive values denote a local excess, and negative values denote a local deficit. The diverging colors use separate linear intervals of $-1000$--$0\,\text{\AA}$ and $0$--$7500\,\text{\AA}$, with zero fixed at white; cell labels give the untransformed EWs. Very large absolute values occur for line-dominated components with little local continuum and should not be interpreted as source-level EWs.}
    \label{fig:agn_diagnostics}
\end{figure*}

To determine what spectral variation the AGN block adds beyond the galaxy components, we examine its smooth continuum shapes and localized emission features in Figure~\ref{fig:agn_diagnostics}. The left panel compares each component with blackbody spectra, while the right panel measures the local equivalent widths (EWs) of ${\rm H}\alpha$, ${\rm Pa}\alpha$, and the PAH $3.3\um$ feature. This PAH band arises from aromatic C--H stretching modes and is often used as a tracer of star formation, although AGN radiation can also excite PAH-bearing clouds in which the molecules remain shielded from destruction \citep{Imanishi2000,Jensen2017}. The galaxy components are included in both panels as a reference for identifying the continuum and emission-feature structure added by the AGN block.

In the left panel, the AGN components reach their maximum blackbody similarity at $T_{\rm BB}=400$--$1200\,\mathrm{K}$, consistent with the warm and hot torus-heated dust that shapes AGN near-infrared continua \citep{Landt2011,Mor2012,Burtscher2015}. By comparison, the galaxy components peak at $3000$--$10{,}000\,\mathrm{K}$, characteristic of the photospheric temperatures contributing strongly to their integrated stellar continua. This separation makes the cooler dust-like continuum structure of the AGN block explicit. Components $A_0$, $A_1$, and $A_3$ have the clearest continuum character, combining relatively high blackbody similarity with weak EWs in all three features. Their weak or negative PAH contrast is consistent with dilution by a strong AGN-heated continuum and with destruction of PAH molecules by hard AGN radiation in nuclear regions \citep{Voit1992,Imanishi2000,AlonsoHerrero2014}.

The right panel reveals complementary feature-dominated structure. Component $A_2$ is dominated by PAH $3.3\um$, with a weaker positive ${\rm Pa}\alpha$ excess, while $A_4$ has very large ${\rm H}\alpha$ and ${\rm Pa}\alpha$ EWs together with strong PAH emission and traces a broader recombination-line complex. These PAH-rich AGN components show that the active-galaxy sample requires PAH-associated spectral variation beyond that represented by the fixed normal-galaxy dictionary. Their training origin does not uniquely identify the exciting source. At the source level, positive reconstructed PAH emission remains present at the high-luminosity end of the AGN sample, while its EW shows no monotonic increase with AGN luminosity. This behavior is consistent with emission from shielded circumnuclear or host-galaxy material, with contributions from stellar and potentially AGN radiation \citep{AlonsoHerrero2014,Jensen2017}. The galaxy components provide a useful reference. Among them, $G_5$ shows the clearest PAH excess but lacks the strong ${\rm H}\alpha$ and ${\rm Pa}\alpha$ emission seen in $A_4$. This difference indicates that the two components trace distinct covariance patterns. Component $G_5$ traces PAH-rich, dusty star formation, consistent with the independent correlations in Section~\ref{ssec:galaxy_validation}, whereas $A_4$ captures a recombination-line-rich complex in the AGN sample. Negative EWs in the other components arise from local continuum deficits that balance the strengths of these features across the non-negative template set. Taken together, the AGN components recover the principal spectral signatures associated with active-galaxy systems, including torus-heated dust continua, PAH emission in shielded material, and strong recombination-line emission.

%% file: 05-discussion.tex
\section{Applications and Limitations}
\label{sec:applications}

We illustrate how the empirical dictionary can support astrophysical inference through three progressive applications. At a fixed spectroscopic redshift, its two component blocks provide an operational separation of host-galaxy and AGN emission. Allowing redshift to vary extends the same representation to redshift estimation. Finally, the fitted coefficients provide a low-dimensional feature space for classifying galaxies and AGNs, extending a traditional use of empirical eigenspectra to an additive, non-negative representation.

\subsection{Near-infrared Host Fractions}
\label{ssec:host_fraction}

To test the AGN--host decomposition, we assemble a sample of 1,778 SDSS quasars with optical spectral decompositions from \citet{Ren2024}, HSC image decompositions from \citet{Li2021}, and SPHEREx coverage. All sources satisfy $m_{\mathrm{AB},W1}<18.5\,\mathrm{mag}$, the same brightness criterion adopted for the training sample, and span $0.2<z<0.8$, as set by the selection criteria of the reference catalogs. Using the available SDSS spectroscopic redshifts, we fix each source at $z_{\rm spec}$ and solve only for the 14 non-negative coefficients, thereby isolating decomposition performance from redshift estimation failures.

For each fit, we separately reconstruct the galaxy and AGN emission from their respective dictionary blocks. We define the near-infrared host fraction, $f_{\rm host}^{\rm IR}$, as the galaxy template flux integrated over rest-frame $2.1$--$2.3\um$ divided by the total flux integrated over the same interval.
Longer wavelengths offer greater sensitivity to AGN-heated dust, but the SPHEREx bandpass restricts direct coverage of such rest-frame wavelengths to progressively lower redshifts. Because most of our sources lie at $z<1$, we adopt $2.1$--$2.3\um$ as the longest practical interval that remains constrained by the observations for the accepted sample. The rest-frame $2.1$--$2.3\um$ interval provides a measurement near the transition between stellar and hot-dust continua.

To ensure adequate infrared coverage for the decomposition, we require at least 204 retained individual photometric measurements across channels and visits, equivalent to two complete sets of the 102 nominal SPHEREx spectral channels. We also mask the observer-frame ${\rm He\,I}\ 1.083\um$ airglow region. Of the 1,778 matched quasars, 721 meet this coverage requirement and constitute the validation sample. Among them, 69 sources lack a valid $f_{\rm host}$ measurement at rest-frame 5100~\AA\ in the \citet{Ren2024} catalog.

\begin{figure*}[ht!]
    \plotone{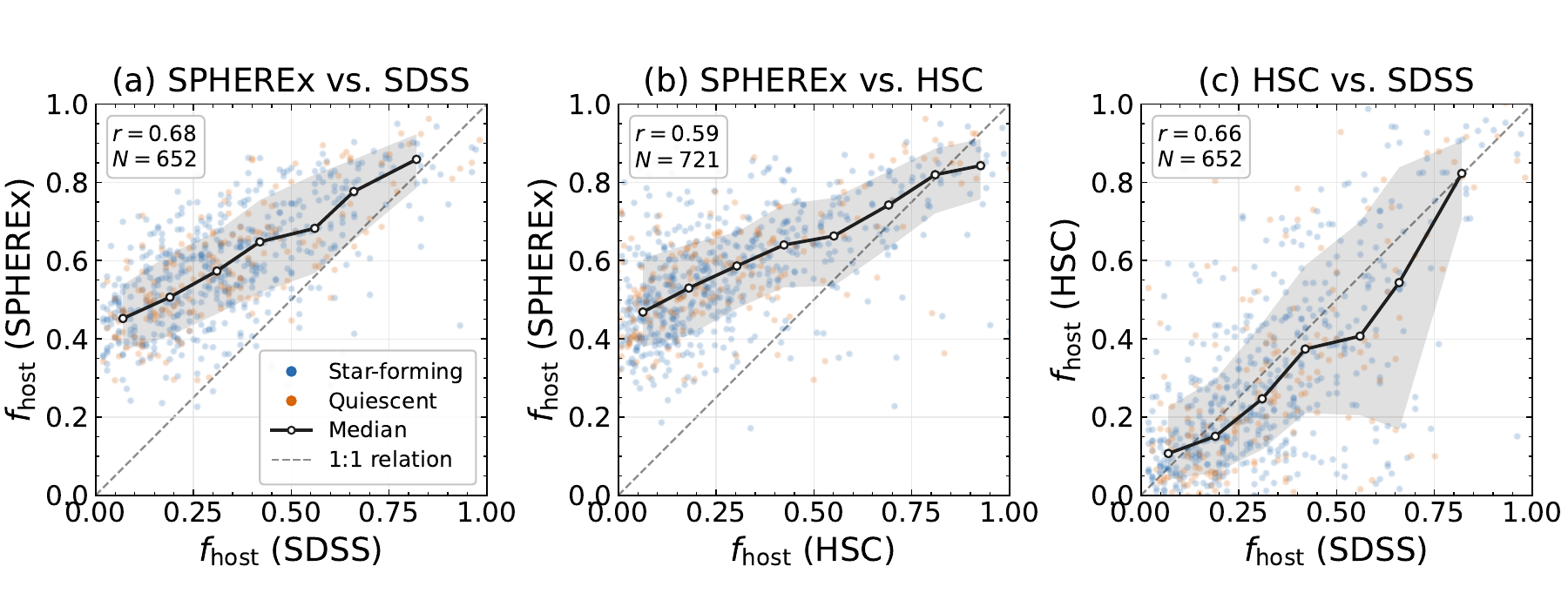}
    \caption{External validation of the near-infrared host fraction. \textit{Left:} $f_{\rm host}$ from the fixed-redshift SPHEREx decomposition over rest-frame $2.1$--$2.3\um$ versus $f_{\rm host}$ at rest-frame 5100~\AA\ from SDSS spectral decomposition \citep{Ren2024}. \textit{Middle:} The SPHEREx host fraction versus $f_{\rm host}$ in the HSC $r$ band from image decomposition \citep{Li2021}. \textit{Right:} The HSC imaging and SDSS spectral measurements compared for their common sample. Blue and orange points denote star-forming and quiescent hosts in the HSC catalog. Black curves and open markers trace the median in bins of the horizontal variable, with gray bands spanning the 16th--84th percentiles; dashed lines mark equality. Pearson $r$ and the number of valid sources are reported in each panel.}
    \label{fig:host_fraction_validation}
\end{figure*}

Figure~\ref{fig:host_fraction_validation} compares the SPHEREx near-infrared host fractions with the SDSS spectral and HSC imaging decompositions. Both comparisons show clear positive correlations. We obtain Pearson $r=0.68$ against the SDSS host fraction at 5100~\AA\ and $r=0.59$ against the HSC $r$-band fraction. For their common sample, the two external measurements correlate at $r=0.66$. The similar strengths of all three correlations place the scatter in the SPHEREx comparisons at the level already present between the two external methods. These results show that the SPHEREx decomposition recovers source-to-source variations in the relative AGN and host contributions, although none of the three methods provides a high-precision reference decomposition. Exact agreement is also limited because the AGN-to-host contrast varies with wavelength, aperture, and decomposition method. In addition, the HSC $r$-band measurement is defined in the observer frame and therefore samples a redshift-dependent rest-frame wavelength.

Unlike the two optical measurements, which extend to host fractions near zero, the $K$-band distribution has a minimum of 0.17 and a fifth percentile of 0.34. The scarcity of similarly low values in the $K$ band is consistent with its strong stellar contribution and the statistical coupling between black hole mass and host stellar mass \citep{Ding2020,Suh2020}. At fixed scaling relation, only unusually high Eddington ratios would drive the $K$-band host fraction toward zero.

The wavelength-sensitivity test further supports this interpretation. Moving the integration interval from $1.0$--$1.2$ to $1.5$--$1.7$, $2.1$--$2.3$, and $2.8$--$3.0\um$ lowers the median host fraction from 0.90 to 0.82, 0.56, and 0.32. The longest-wavelength interval reaches a minimum of 0.07 as the stellar continuum declines and hot dust becomes more prominent. This wavelength dependence supports a primarily physical origin for the scarcity of near-zero $K$-band host fractions. A further, likely minor contribution may arise from the operational dictionary split. Smooth continuum structure shared by normal galaxies and AGN is already represented by the frozen galaxy block, so part of an AGN continuum with a similar shape may be assigned to that block. Non-negativity prevents the remaining components from correcting such an allocation through subtraction, potentially biasing the lowest inferred host fractions upward.

\subsection{Redshift Estimation}
\label{ssec:redshift_determination}

The preceding host-fraction analysis fixes the spectroscopic redshift to isolate the AGN--host decomposition. Applying the dictionary without external spectroscopy additionally requires the redshift to be inferred from the same SPHEREx measurements. We therefore fit directly in the observer frame, allowing the redshift and component weights to vary jointly. For each trial redshift $z$, the rest-frame dictionary is shifted on a logarithmic wavelength grid and projected through the source response matrix. The coefficients are obtained by NNLS, and the solution is selected by
\begin{equation}
\begin{aligned}
    (z_{\rm fit},\mathbf{w}_{\rm fit})
    &= \arg\min_{z,\,\mathbf{w}\geq0}\chi^2(z,\mathbf{w}), \\
    \chi^2(z,\mathbf{w})
    &= \sum_{i\in\mathcal{C}(z)}
    \left[
    \frac{y_i-[\mathbf{R}_{\rm obs}\mathbf{V}(z)\mathbf{w}]_i}
    {\sigma_i}
    \right]^2 \\
    &\quad + \sum_{i\notin\mathcal{C}(z)}
    \left(\frac{y_i-c_i}{\sigma_i}\right)^2.
\end{aligned}
    \label{eq:redshift_fit}
\end{equation}
Here, $y_i$ and $\sigma_i$ are the observed flux and its uncertainty, $\mathbf{R}_{\rm obs}$ is the source response matrix, $\mathbf{V}(z)$ contains the shifted templates, and $\mathbf{w}$ contains the non-negative component weights. The set $\mathcal{C}(z)$ contains observations whose response windows are fully covered by the shifted templates. Measurements outside $\mathcal{C}(z)$ are excluded from the NNLS coefficient fit but remain in the total $\chi^2$ through $c_i$, the value of a fixed weighted linear baseline fitted separately in each detector band. This deliberately simple baseline retains the residual scatter of the uncovered measurements in $\chi^2$. Redshift trials with different template coverage are therefore evaluated using the same set of observations, preventing high-redshift solutions from gaining an artificial advantage by shifting the templates away from high-S/N measurements.

\begin{figure*}[ht!]
    \plotone{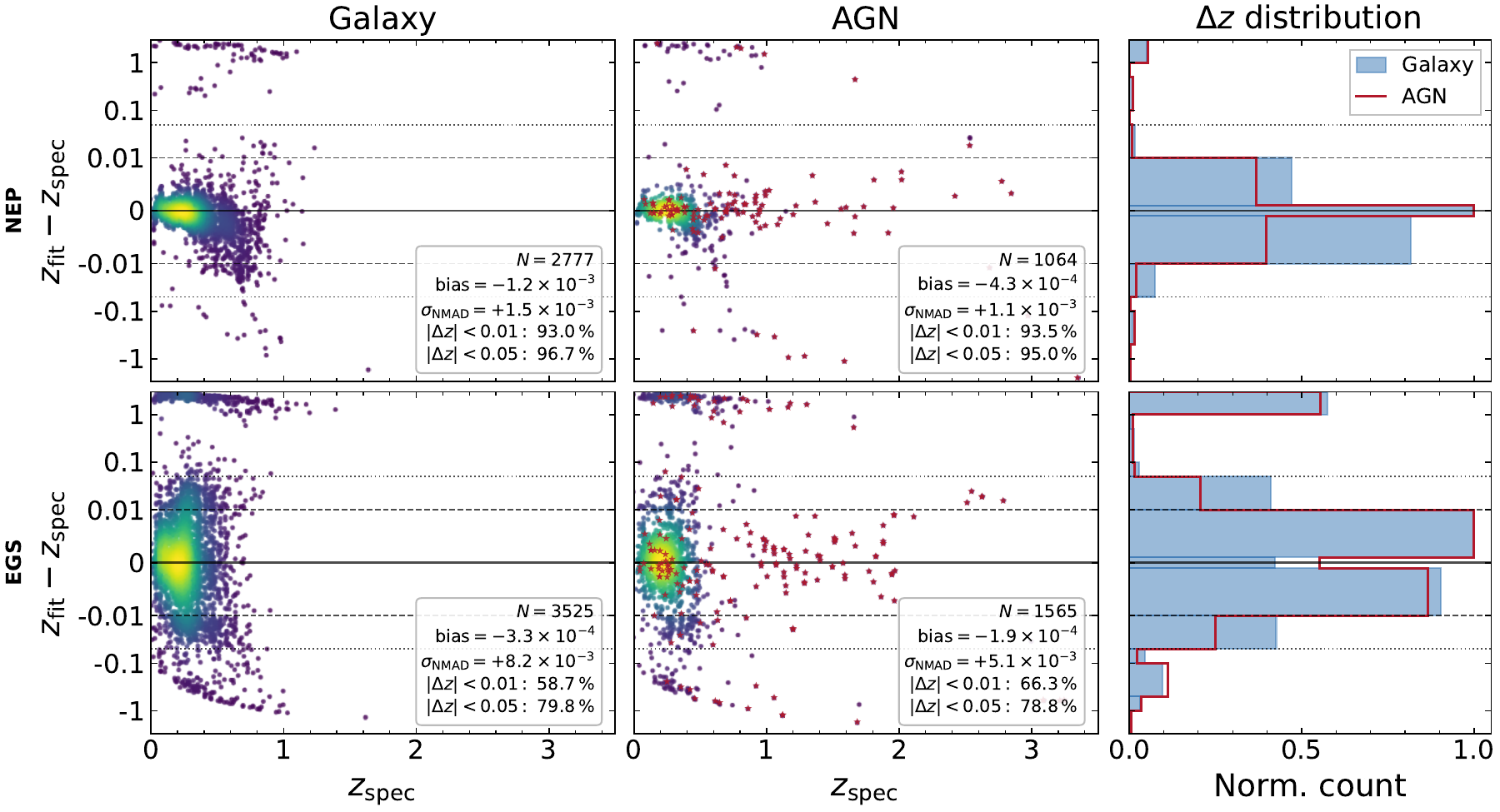}
    \caption{Template-fitting redshift performance for normal galaxies and AGN in the NEP deep field (top) and EGS shallow field (bottom). The left and middle columns show $\Delta z=z_{\rm fit}-z_{\rm spec}$ as a function of DESI spectroscopic redshift on a symmetric-logarithmic vertical scale. Red triangles identify broad-line AGN. Solid, dashed, and dotted lines mark $\Delta z=0$, $|\Delta z|=0.01$, and $|\Delta z|=0.05$. Insets report the sample size, clean-sample bias, normalized median absolute deviation, and fractions within the two absolute-error thresholds. The right column compares peak-normalized $\Delta z$ distributions for galaxies and AGN. Both fields use the same unrestricted search range, $-0.1\leq z\leq3.0$.}
    \label{fig:redshift_validation}
\end{figure*}

The observer-frame grid has the same logarithmic spacing as the template grid, so each redshift trial is implemented as an integer shift without interpolating the templates. To reduce the computational cost, we use a two-stage search over $-0.1\leq z\leq3.0$: a coarse scan with $\Delta z_{\rm grid}\simeq0.01$ identifies the three best local minima, each of which is refined at $\Delta z_{\rm grid}\simeq0.001$. We define $\Delta z=z_{\rm fit}-z_{\rm spec}$, $\delta z=\Delta z/(1+z_{\rm spec})$, and report $\sigma_{\rm NMAD}=1.4826\,{\rm median}|\delta z-{\rm median}(\delta z)|$. The bias shown in Figure~\ref{fig:redshift_validation} is the mean $\Delta z$ after excluding $|\Delta z|\geq0.1$.

Both the NEP and the EGS results show narrower AGN redshift errors than galaxy redshift errors, consistent with strong emission lines providing localized anchors. Because the AGN-trained templates are learned only from active sources but applied to all objects, we test whether they introduce spurious galaxy solutions. Using only the seven galaxy templates changes the galaxy catastrophic fraction from $3.13\%$ to $3.24\%$, but raises the AGN fraction from $4.79\%$ to $15.04\%$. The AGN-trained templates therefore add population-specific redshift information without degrading the galaxy fits.

The shallower EGS data broaden the error distributions to $\sigma_{\rm NMAD}=8.19\times10^{-3}$ for galaxies and $5.12\times10^{-3}$ for AGN. These values remain below the normalized scatters of 0.039 and 0.110 reported by \citet{Jiang2026} for galaxy and quasar photometric redshifts from nine-band imaging. Although their differing redshift distributions preclude a direct comparison, this improvement is expected from narrow-band spectrophotometry. A dominant cause of the relatively high catastrophic fraction in EGS is the matching of Paschen-series features to Balmer lines, which assigns spuriously high redshifts to low-redshift sources. Restricting the fit to $z\leq1.5$ increases the fraction with $|\Delta z|<0.05$ from $79.5\%$ to $89.0\%$ and reduces the catastrophic fraction from $18.9\%$ to $8.4\%$. Given the SPHEREx limiting magnitude at all-sky depth, this restriction is feasible for much of the detectable source population.

From another perspective, we can exploit atypical component mixtures to identify unreliable redshift solutions. We therefore train a random forest \citep{Breiman2001} on the normalized weights, $z_{\rm fit}$, and $\chi^2$. In five-fold NEP out-of-fold predictions, we find that the lowest-scoring $5\%$ contain 125 of the 138 catastrophic failures ($90.6\%$), allowing us to flag most failures even when the correct redshift cannot be determined. These experiments support the potential for redshift inference for bright sources across the SPHEREx all-sky survey. However, a calibrated all-sky redshift catalog is beyond our scope. We also note that performance will degrade beyond the present limit of $W1<18.5$.

\subsection{Classification in the Component-weight Space}
\label{ssec:classification}

The fitted coefficients also provide a low-dimensional representation for source classification. We define three operational classes following Section~\ref{sec:data}: BLAGN/WISE combines sources selected through broad lines or WISE colors, NLAGN/other contains sources identified by the remaining AGN diagnostics, and Galaxy denotes sources without an AGN flag. Every source in this validation sample has DESI spectroscopy, which provides both the reference classification and the spectroscopic redshift $z_{\rm spec}$. We fix each source at $z_{\rm spec}$ solely to measure classification from the near-infrared coefficients without including redshift-estimation errors. This validation choice does not require spectroscopy when applying the classifier to additional sources: the required redshift can instead be estimated from SPHEREx together with complementary multiband data. Forecasts for bright galaxies reach $\sigma_{\rm NMAD}\lesssim0.005$, substantially better than conventional broadband-only photometric redshifts \citep{Bae2026}.

The classifier uses 16 features: the 13 ratios $w_k/w_0$ for $k=1$--$13$, the integrated template fluxes $f_{\rm gal}$ and $f_{\rm AGN}$, and $z_{\rm spec}$. We compute these fluxes from the weighted galaxy and AGN-trained templates over the full common template wavelength range, ensuring comparability across redshift.

\begin{figure*}[ht!]
    \plotone{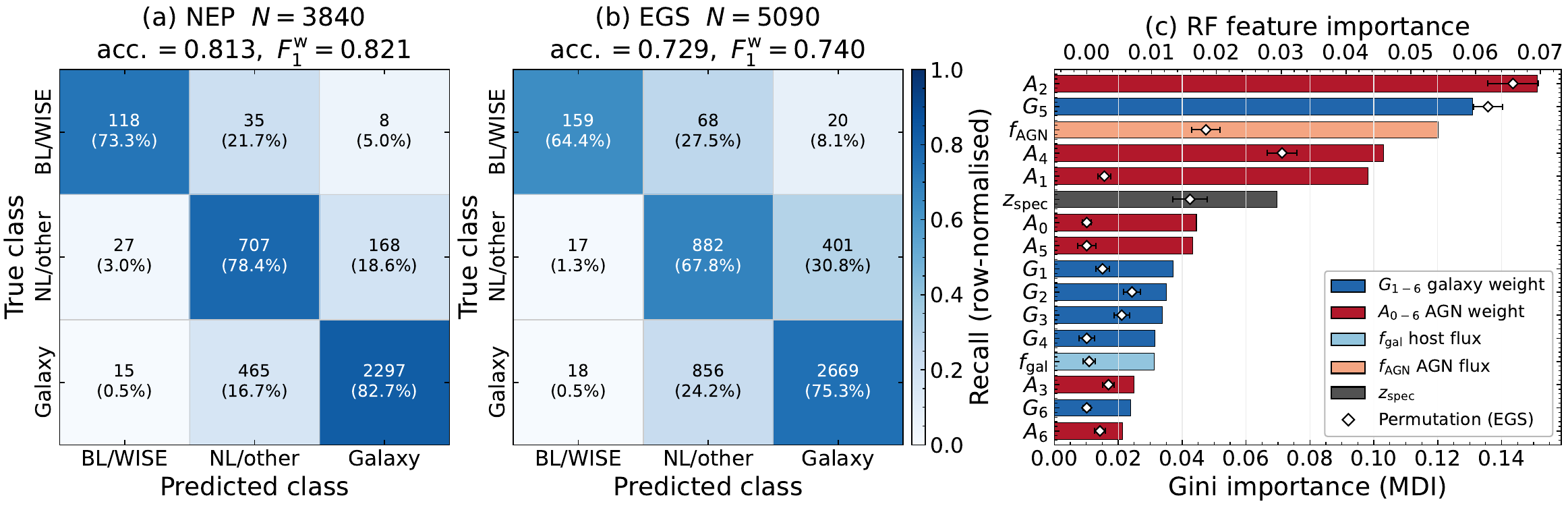}
    \caption{Three-class classification and feature importance from coefficients fitted at $z_{\rm spec}$. Panel (a) shows the five-fold cross-validation results in NEP, while panel (b) evaluates the full-NEP model on the shallower EGS data. Rows give the true class, columns give the predicted class, and percentages give the row-normalized recall. Panel (c) ranks the 16 inputs by the mean decrease in impurity (MDI). Diamonds show the decrease in weighted $F_1$ over 10 EGS permutations.}
    \label{fig:classification}
\end{figure*}

For a representative result, we use a simple 200-tree random forest \citep{Breiman2001}. We follow nested model-selection practice to keep hyperparameter tuning separate from performance evaluation \citep{Varma2006,Cawley2010}. We divide the NEP sample into five subsets with similar class proportions, train on four, and predict the held-out fifth, repeating the process until every source has been predicted once. Within each four-subset training sample, an additional four-fold search selects the tree depth, minimum leaf size, number of candidate features per split, and class weights that maximize the weighted $F_1$. The five held-out prediction sets are then combined to form the NEP confusion matrix (left panel of Figure~\ref{fig:classification}). For the EGS test, all model selection remains confined to NEP. Five-fold cross-validation on the full NEP sample selects the adopted configuration: a maximum depth of 15, a minimum of three samples per leaf, $\sqrt{N_{\rm feature}}$ candidate features per split, and class weights of 1:3:5 for Galaxy, NLAGN/other, and BLAGN/WISE. We then fit one model to all NEP sources and apply it to EGS without additional adjustment.

We quantify feature importance in two complementary ways. The mean decrease in impurity (MDI) sums how strongly each feature reduces class mixing across the trees in the final full-NEP model. We also permute each feature independently in EGS and measure the resulting decrease in weighted $F_1$. The diamonds in Figure~\ref{fig:classification} show the mean decrease over 10 permutations, with error bars indicating the standard deviation.

The NEP cross-validation gives an accuracy of $81.3\%$ and weighted $F_1=0.821$ (Figure~\ref{fig:classification}). These metrics decrease to $72.9\%$ and 0.740, respectively, in EGS. Because both fields are fitted at $z_{\rm spec}$, this decline reflects the weaker constraints on the template coefficients from the shallower all-sky-like data rather than catastrophic redshift failures. The main loss occurs between NLAGN/other and Galaxy. These classes are intrinsically difficult to separate because AGN lacking broad lines or WISE-selected torus emission are generally host dominated. With shallower sampling, their weaker emission features are less well reconstructed by templates, further suppressing the classification signal. The feature-importance analysis below also supports this interpretation.

The feature ranking identifies emission-line structure as the primary classification signal. The line-rich components $A_2$, $G_5$, and $A_4$ have the highest MDI importance among the template weights (Figure~\ref{fig:agn_diagnostics}). Consistently, permuting $A_2$ or $G_5$ in EGS produces the largest performance losses. The shallower EGS sampling weakens the constraints on these components, making host-dominated AGNs more readily confused with galaxies. Continuum information remains complementary. Although the high-color-temperature components $A_0$ and $A_1$ trace hot-torus emission and rank highly in MDI, their lower EGS permutation importance indicates limited discrimination in the shallower data.

The apparent classification errors also expose limitations in the ground-truth labels. Although DESI provides one of the most advanced spectroscopic data sets available, its catalog classifications and redshifts are not infallible. Among NEP sources predicted as BLAGN/WISE but labeled NLAGN/other or Galaxy, visual inspection identifies three spectra with clear broad ${\rm H}\alpha$ and three with incorrect spectroscopic redshifts, together accounting for $14\%$ of this error subset. The corresponding EGS inspection finds five broad-${\rm H}\alpha$ sources and three redshift failures ($23\%$). These cases are more readily recognized within this particular error channel and therefore do not measure the overall DESI label-error rate. Nevertheless, their concentration among the nominal false positives shows that the near-infrared coefficients can independently flag questionable optical classifications. Label ambiguity is likely still greater between NLAGN/other and Galaxy. The stacked SPHEREx SEDs of the four confusion categories show that galaxy-labeled sources predicted as NLAGN/other have, on average, slightly stronger ${\rm Pa}\alpha$ emission and a near-infrared torus excess than correctly classified galaxies, although the source-to-source scatter remains large (Appendix~\ref{sec:app_classification_confusion}). Thus, part of the apparent confusion may reflect weak AGN activity missing from the adopted ground truth rather than classifier failure alone.

\subsection{Limitations}
\label{ssec:limitations}

The present templates are directly supported for SPHEREx spectrophotometry of bright ($m_{\mathrm{AB},W1}<18.5\,\mathrm{mag}$), predominantly low-redshift galaxies and AGNs within their rest-frame $0.45$--$4.60\um$ coverage. Their transfer from the NEP training field to the shallower EGS data without retraining shows that the learned spectral basis remains informative at all-sky-like depth, although measurement depth and wavelength coverage affect the precision of the fitted coefficients and downstream inferences. Fainter, higher-redshift, or differently selected samples may contain galaxy and AGN populations not represented in the present training set, so transfer to these regimes requires validation with representative data. The rest-frame components can also be projected through other instrumental responses. For data at higher spectral resolution than SPHEREx, such validation should first match the data to SPHEREx-like resolution because the effective resolving power of the reconstructed components has not been independently established.

The most important limitation within this domain is the lack of complete ground-truth labels for AGN activity. Our two-stage training assumes that the adopted diagnostics identify AGNs with sufficient completeness to define a clean normal-galaxy sample, but highly obscured or weak, host-dominated nuclei are easily missed \citep{Hickox2018}. Such sources would contaminate the galaxy training sample and imprint AGN-related structure on the fixed galaxy templates, potentially limiting the separation between the two blocks. An additional source of label mismatch is the difference in effective aperture. Most optical emission-line diagnostics in the DESI VAC are measured from $1\farcs5$-diameter fiber spectra and therefore preferentially sample the central regions of the galaxies, whereas our SPHEREx photometry is measured in apertures with diameters of $2.5\times\mathrm{FWHM_{PSF}}$ and therefore includes a much larger fraction of the host galaxy. Nuclear activity identified from the DESI spectrum can consequently be diluted by integrated stellar and star-forming emission in the SPHEREx SED. Independent X-ray information would provide valuable complementary labels for such activity. Although ROSAT provides deeper X-ray coverage in the NEP field, its sensitivity is only sufficient for X-ray-bright AGNs and is inadequate for a representative cross-check of our sample \citep{Henry2006}. The public eROSITA eRASS1 release covers only the western Galactic hemisphere and therefore does not include the NEP field \citep{Merloni2024}. We leave a systematic X-ray validation to future public releases or deeper pointed coverage.

Additional limitations arise from the current SPHEREx measurements. Because per-pixel transmission curves were unavailable, we approximate each passband as a top-hat, which may affect the effective resolution of the recovered templates. Although torus emission varies on long timescales \citep{Li2023}, its variability is not negligible over the span of the SPHEREx visits. The higher AGN-stage $\chi^2/M_{\rm tot}$ of 1.34, compared with 1.05 for galaxies, indicates additional scatter that can blur the recovered templates when multi-epoch measurements are combined.

%% file: 06-conclusion.tex
\section{Summary}
\label{sec:conclusion}

We have constructed an empirical near-infrared spectral dictionary directly from SPHEREx spectrophotometry while accounting for the position-dependent wavelength coverage and passband response of the linear variable filters. The framework learns rest-frame templates on a common $0.45$--$4.60\um$ grid and projects their non-negative combinations into the observed flux space using a source-specific response matrix. Sequential training on 2,777 normal galaxies and 1,064 sources with AGN diagnostic flags yields a $K=14$ dictionary comprising seven galaxy and seven AGN-trained components.

This construction separates spectral variation shared with normal galaxies from the additional variation required by the AGN sample. The galaxy block spans old-stellar and dusty, line-rich spectral modes that track independently measured stellar-population properties and reproduce E-MILES continua with a median coverage-weighted residual of $2.1\%$. The AGN block adds $400$--$1200\,\mathrm{K}$ dust-like continua and prominent ${\rm H}\alpha$, ${\rm Pa}\alpha$, and PAH $3.3\um$ structure associated with active-galaxy environments.

We tested the information retained by this representation through three applications:
\begin{enumerate}
    \item \textit{Host--AGN decomposition.} For 721 quasars, the inferred rest-frame $2.1$--$2.3\um$ host fractions correlate with independent SDSS spectral and HSC imaging decompositions at Pearson $r=0.68$ and $r=0.59$, respectively. The median host fraction decreases from 0.90 at $1.0$--$1.2\um$ to 0.32 at $2.8$--$3.0\um$, tracing the transition from stellar to hot-dust emission.
    \item \textit{Redshift estimation.} Dictionary fitting yields $|\Delta z|<0.01$ for $93.0\%$ of galaxies and $93.5\%$ of AGNs in NEP, with normalized component weights identifying $90\%$ of catastrophic failures.
    \item \textit{Source classification.} At fixed $z_{\rm spec}$, the fitted coefficients classify galaxies, BLAGNs, and other AGNs with accuracies of $81.3\%$ in NEP and $72.9\%$ in the shallower EGS field.
\end{enumerate}

The resulting dictionary provides a common empirical representation for host--AGN decomposition in large SPHEREx samples and a basis for combining these decompositions with complementary optical and imaging measurements. The framework is designed for survey-scale application and can grow with the available data volume. Together, these capabilities enable robust population-level investigations of galaxy populations, black hole activity, and their coevolution. As SPHEREx calibration improves and larger multiwavelength reference samples become available, the dictionary can be retrained within the same framework to refine the training labels and broaden the populations represented.

The analysis code and empirical spectral templates presented in this work are archived on Zenodo \citep{ren_2026_22226283}.

%% file: appendix.tex
\section{Overcomplete ALS-WNMF Exploration}
\label{sec:app_optimization}

The specific discretized forms of the regularization terms $\mathcal{R}(\mathbf{V})$ used to enforce template stability and smoothness are:
\begin{align}
    \mathcal{R}_{L2}(\mathbf{v}_k) &= \frac{1}{T} \sum_{t=1}^{T} V_{tk}^2, \\
    \mathcal{R}_{\mathrm{diff}}(\mathbf{v}_k) &= \frac{1}{T-1} \sum_{t=1}^{T-1} (V_{t+1, k} - V_{tk})^2, \\
    \mathcal{R}_{\mathrm{curv}}(\mathbf{v}_k) &= \frac{1}{T-2} \sum_{t=2}^{T-1} (V_{t+1, k} - 2V_{tk} + V_{t-1, k})^2,
\end{align}
where $\mathbf{v}_k \in \mathbb{R}^T$ represents the $k$-th column of $\mathbf{V}$. The parameters $\alpha$, $\beta$, and $\gamma$ control the strengths of the $L_2$, first-order smoothness, and second-order curvature penalties, respectively.

Both the galaxy and AGN training stages begin with an overcomplete ALS-WNMF exploration based on alternating non-negative coefficient and template updates \citep{Lee2001,Blanton2007}. In the galaxy stage, all 40 candidate templates are trainable. In the AGN stage, the dictionary is partitioned as $\mathbf{V}=[\mathbf{V}^{\mathrm{fix}}\mid\mathbf{V}^{\mathrm{free}}]$. The seven galaxy templates in $\mathbf{V}^{\mathrm{fix}}$ remain frozen, while the 25 candidate templates in $\mathbf{V}^{\mathrm{free}}$ are trainable. Each exploratory fit alternately updates the source coefficients and the trainable template columns.
\begin{itemize}
    \item \textbf{Coefficient update:} Holding the basis matrix $\mathbf{V}$ fixed, the coefficient vector $\mathbf{w}_n$ for each source $n$ is optimized independently via NNLS:
    \begin{equation}
        \mathbf{w}_n = \arg\min_{\mathbf{w} \ge 0} \left( \mathbf{y}_n - \mathbf{R}_n \mathbf{V} \mathbf{w} \right)^{\mathsf{T}} \mathbf{\Sigma}_n^{-1} \left( \mathbf{y}_n - \mathbf{R}_n \mathbf{V} \mathbf{w} \right).
    \end{equation}
    \item \textbf{Template update:} Holding the coefficients $\mathbf{W}$ fixed, all trainable columns are updated simultaneously with the same multiplicative update rule. This means updating the full dictionary in the galaxy exploration and only $\mathbf{V}^{\mathrm{free}}$ in the AGN exploration. The frozen columns $\mathbf{V}^{\mathrm{fix}}$ still contribute to the AGN reconstruction and coefficient fit but are excluded from the template update.
\end{itemize}

\section{Implementation Details of Similarity-Based Pruning and Merging}
\label{sec:app_prune}

Initial over-parameterized training yields redundant templates that span similar spectral features. To compress the dictionary and resolve redundancy, we perform similarity-based pruning:
\begin{enumerate}
    \item \textbf{Smoothing:} Each candidate basis vector $\mathbf{v}_i$ is smoothed along the wavelength axis using a 1D Gaussian kernel with width $\sigma=5~\mathrm{pixels}$ to prevent high-frequency noise from dominating the similarity metric (referred to as proxy smoothing).
    \item \textbf{Cosine Similarity:} We compute the pairwise cosine similarity matrix $\mathbf{S} \in \mathbb{R}^{K \times K}$:
    \begin{equation}
        S_{ij} = \frac{\mathbf{v}_i \cdot \mathbf{v}_j}{\| \mathbf{v}_i \|_2 \, \| \mathbf{v}_j \|_2}.
    \end{equation}
    \item \textbf{Hierarchical Clustering:} We convert the similarities to distances ($d_{ij} = 1 - S_{ij}$) and perform agglomerative hierarchical clustering with complete linkage. Components belonging to clusters with a distance smaller than $1 - \tau$ (where $\tau$ is the cosine similarity threshold) are grouped together.
    \item \textbf{Weighted Merging:} Redundant components in each cluster $\mathcal{C}$ are merged. The merged template shape is computed as a weighted average using their global weights:
    \begin{equation}
        \mathbf{v}_{\mathrm{merged}} = \frac{\sum_{i \in \mathcal{C}} a_i \, \mathbf{v}_i}{\sum_{i \in \mathcal{C}} a_i},
    \end{equation}
    where $a_i = \sum_{n=1}^N W_{ni}$ represents the cumulative contribution of component $i$ across the sample. The corresponding weights are summed to preserve the total activation:
    \begin{equation}
        W_{n, \mathrm{merged}} = \sum_{i \in \mathcal{C}} W_{ni}.
    \end{equation}
    \item \textbf{Ranking and Truncation:} The remaining templates are sorted in descending order of their global contribution $a_i$, and the dictionary is truncated to the desired dimension.
\end{enumerate}
Before HALS refinement, we rescale each retained trainable template to unit mean and apply the compensating scale change to its coefficients for numerical conditioning; this leaves the reconstructed spectra unchanged.

\section{Selection of the Number of Templates}
\label{sec:app_dictionary_size}

We select the number of templates by balancing underfitting against redundancy. With too few templates, distinct spectral shapes recurring across the sample are forced into the same component, leaving predictable signal unmodeled. With too many, similar continuum or emission-line structure can be divided among nearly interchangeable components. The fit then improves slightly, but the inferred mixture becomes sensitive to the particular measurements used. The monotonically decreasing training loss alone cannot distinguish these regimes.

We therefore use the repeated SPHEREx measurements to test whether a candidate dictionary predicts independent observations of the same source. Within each detector band and wavelength cell, MJD-ordered visits are assigned alternately to two halves so that both retain comparable wavelength coverage. Coefficients fitted to one half are used to predict the other, giving a cross-half $\chi^2$ per measurement. We also measure how much normalized coefficient weight moves between the two fits,
\begin{equation}
    D_{w,n} = \frac{1}{2}\sum_k \left|f_{nk}^{(1)}-f_{nk}^{(2)}\right|,
\end{equation}
where $D_{w,n}=0$ denotes identical component allocation. A useful additional template should therefore lower the cross-half loss, recur in a non-negligible fraction of the sample, and avoid a large increase in $D_w$.

Candidate dictionaries at each exact rank are drawn from the pruning hierarchy. After removing the lowest-participation $20\%$ of the overcomplete components, we cluster Gaussian-smoothed template proxies with complete linkage and refit the coefficients with NNLS. A component is counted as active in a source when its normalized coefficient satisfies $f_{nk}>0.01$. We retain sources with at least 40 valid measurements, coverage in at least four detector bands, and a valid visit-parity split, yielding 2,138 normal galaxies and 819 AGN.

\begin{figure*}[ht!]
    \plotone{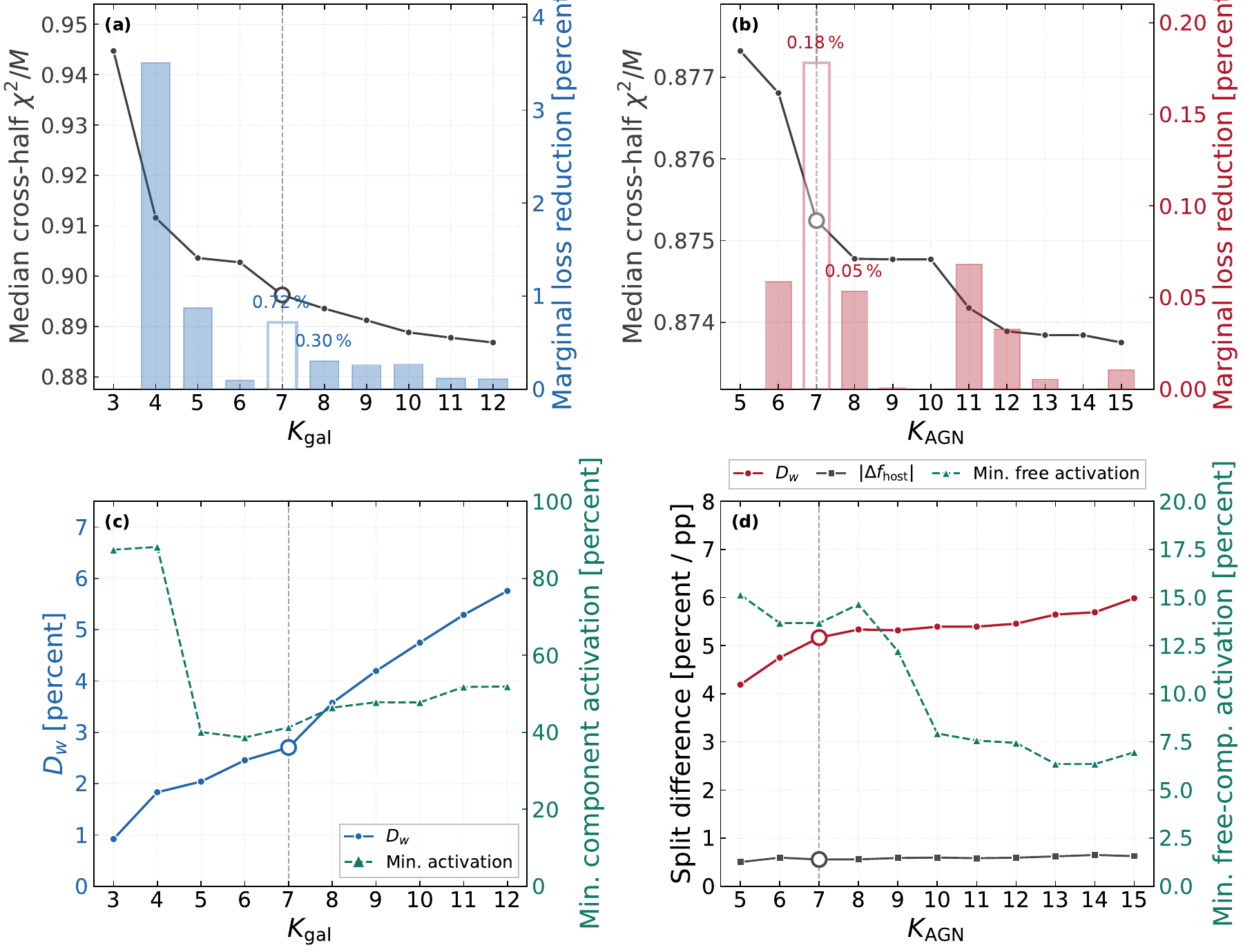}
    \caption{Dictionary-rank selection diagnostics evaluated before HALS refinement. In the top panels, the black curves and left axes show the median cross-half $\chi^2$ per measurement, while the colored bars and right axes show its marginal reduction when one component is added to the galaxy dictionary (\textit{a}) or to the AGN-trained free block conditional on seven fixed galaxy templates (\textit{b}). The bottom panels compare the median reassigned coefficient mass, $D_w$, with the minimum component activation fraction for the galaxy candidates (\textit{c}); the corresponding AGN quantities are shown together with the median absolute change in the aggregate host fraction between the two visit halves (\textit{d}). Repeated observations of each source are divided by a band-preserving visit-parity split, and coefficients fitted to either half are used to predict the other. Dashed vertical lines and open markers indicate the adopted rank of seven in each dictionary block. The diagnostics use 2,138 normal galaxies and 819 AGN satisfying the measurement and wavelength-coverage criteria.}
    \label{fig:app_rank_determination}
\end{figure*}

For the galaxy dictionary, we evaluate $K_{\mathrm{gal}}=3$--12. The steep decrease in cross-half loss at low rank shows that the first few templates recover spectral diversity shared by many galaxies. Adding the seventh component provides a further $0.72\%$ reduction, whereas the eighth yields only $0.30\%$ and subsequent additions bring similarly small gains. At the same time, coefficient reassignment increases with rank, indicating that the extra flexibility is increasingly expressed by redistributing flux among components rather than by improving predictions of unseen visits. All seven retained components are active in more than $5\%$ of the eligible galaxies, so the selected rank is not driven by a small number of unusual objects. We therefore adopt $K_{\mathrm{gal}}=7$.

For the AGN sample, the seven fixed galaxy templates first account for spectral structure also present in normal galaxies; the free block then captures the additional diversity required by the AGN population. We evaluate $K_{\mathrm{AGN}}=5$--15 for this block. The seventh free component reduces the cross-half loss by $0.18\%$, but the eighth adds only $0.05\%$. At $K_{\mathrm{AGN}}=7$, every free component occurs in more than $5\%$ of the eligible AGN, and the aggregate fraction assigned to the fixed galaxy block changes by only $0.56$ percentage points between the two visit halves. Thus, the host--AGN allocation remains stable while the predictive improvement has reached a plateau. We adopt $K_{\mathrm{AGN}}=7$, giving a final dictionary of seven fixed galaxy templates and seven AGN-trained templates.

\section{HALS Refinement of the Pruned Dictionaries}
\label{sec:app_hals}

While standard WNMF is effective for broad exploration, it often suffers from the ``twin spectra'' problem, in which multiple templates become highly correlated and duplicate features due to the symmetry of simultaneous updates. After similarity-based pruning, we break this symmetry and accelerate convergence by refining each compact dictionary with a HALS-style block-coordinate algorithm that updates each trainable column sequentially against the residuals left by the other templates \citep{Cichocki2007}. The resulting non-negative single-column subproblem is solved with a diagonally preconditioned projected gradient descent (PGD) inner loop rather than a closed-form HALS update \citep{Lin2007}.

Mathematically, for each trainable component $k$, we update $\mathbf{v}_k$ while keeping all other components $\mathbf{v}_j$ ($j \ne k$) and the coefficients $\mathbf{W}$ fixed. The gradient of the data-reconstruction term with respect to $\mathbf{v}_k$ is:
\begin{equation}
    \frac{\partial \mathcal{L}_{\mathrm{data}}}{\partial \mathbf{v}_k} = 2 \sum_{n=1}^N w_{nk} \mathbf{R}_n^{\mathsf{T}} \mathbf{\Sigma}_n^{-1} \left( \mathbf{R}_n \mathbf{V} \mathbf{w}_n - \mathbf{y}_n \right).
\end{equation}
To make this computationally feasible for large samples, we precompute the first-order constants $\mathbf{C} \in \mathbb{R}^{N \times T}$ and the second-order response matrices $\mathbf{B}_n \in \mathbb{R}^{T \times T}$:
\begin{align}
    \mathbf{C}_{n, :} &= \mathbf{R}_n^{\mathsf{T}} \mathbf{\Sigma}_n^{-1} \mathbf{y}_n, \\
    \mathbf{B}_n &= \mathbf{R}_n^{\mathsf{T}} \mathbf{\Sigma}_n^{-1} \mathbf{R}_n.
\end{align}
We define the aggregation tensor $\mathbf{M} \in \mathbb{R}^{K \times K \times T \times T}$ through blocks $\mathbf{M}_{jk} \in \mathbb{R}^{T \times T}$:
\begin{equation}
    \mathbf{M}_{jk} = \sum_{n=1}^N w_{nj} w_{nk} \mathbf{B}_n,
\end{equation}
and the projection matrix $\mathbf{P} = \mathbf{C}^{\mathsf{T}} \mathbf{W} \in \mathbb{R}^{T \times K}$, the gradient simplifies to:
\begin{equation}
    \frac{\partial \mathcal{L}_{\mathrm{data}}}{\partial \mathbf{v}_k} = 2 \left( \mathbf{M}_{kk} \mathbf{v}_k + \sum_{j \ne k} \mathbf{M}_{jk} \mathbf{v}_j - \mathbf{P}_{:, k} \right).
\end{equation}
We isolate the residual target vector $\mathbf{U}_k \in \mathbb{R}^T$:
\begin{equation}
    \mathbf{U}_k = \mathbf{P}_{:, k} - \sum_{j \ne k} \mathbf{M}_{jk} \mathbf{v}_j.
\end{equation}
The optimization for the $k$-th template then reduces to
\begin{equation}
    \min_{\mathbf{v}_k \ge 0} \mathbf{v}_k^{\mathsf{T}} \mathbf{M}_{kk} \mathbf{v}_k - 2 \mathbf{v}_k^{\mathsf{T}} \mathbf{U}_k + \mathcal{R}(\mathbf{v}_k).
\end{equation}
Using a gradient convention in which the common factor of two is removed, we solve this single-column optimization with five diagonally preconditioned PGD iterations:
\begin{equation}
    \mathbf{v}_k \leftarrow \max \left( 0, \, \mathbf{v}_k - \mathbf{D}_{k}^{-1} \nabla_{\mathbf{v}_k} \mathcal{L} \right).
\end{equation}
Here $\mathbf{D}_k=\operatorname{diag}(\mathbf{d}_k)$ is a positive diagonal majorizer. Its data contribution is constructed from the row sums of the non-negative matrix $\mathbf{M}_{kk}$. Because the $L_2$, first-difference, and curvature penalties are quadratic, their second derivatives are fixed matrices; the absolute row sums of these matrices provide corresponding conservative diagonal bounds. This construction gives stable element-wise step sizes without forming or inverting a dense matrix. The vector $\nabla_{\mathbf{v}_k}\mathcal{L}$ denotes the corresponding regularized gradient.

\section{Spectral Characteristics of Classification-Confusion Subsamples}
\label{sec:app_classification_confusion}

The dominant ambiguity in the three-class experiment of Section~\ref{ssec:classification} occurs between NLAGN/other and Galaxy. To examine the spectral character of this ambiguity, we extract the corresponding $2\times2$ submatrix after excluding every source whose true or predicted class is BLAGN/WISE. We denote correctly classified NLAGN/other and Galaxy sources as TP and TN, respectively. FP denotes galaxies classified as NLAGN/other, and FN denotes NLAGN/other sources classified as galaxies. The cell assignments use the same NEP out-of-fold and EGS transfer predictions as Figure~\ref{fig:classification}.

For each source, we shift the quality-controlled SPHEREx observations to the rest frame using $z_{\rm spec}$ and normalize them by the fitted total SED evaluated at $1\um$. We then assign the individual observations to the common rest-frame wavelength grid. Within each wavelength bin, every covered source receives unit total weight, independent of its number of visits, and the stack is summarized by the weighted median and 16th--84th percentiles. Bins covered by fewer than five independent sources are omitted. This weighting prevents sources with more visits from disproportionately influencing the stack.

\begin{figure*}[ht!]
    \plotone{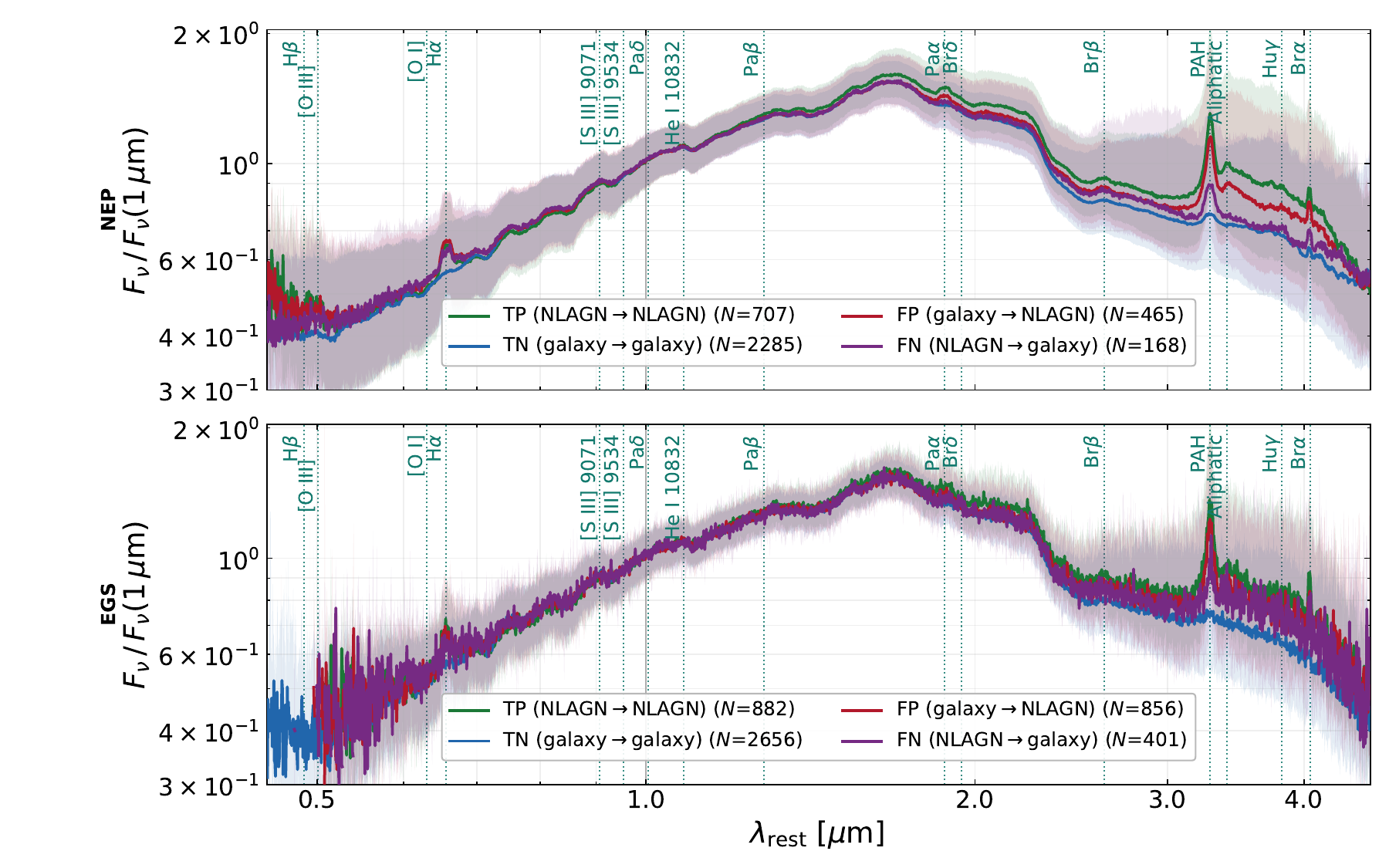}
    \caption{Source-balanced SPHEREx SED stacks for the NLAGN/other--Galaxy confusion cells in NEP (top) and EGS (bottom). TP and TN denote correctly classified NLAGN/other and Galaxy sources, respectively. FP denotes galaxies classified as NLAGN/other, whereas FN denotes NLAGN/other sources classified as galaxies. Sources with either a true or predicted BLAGN/WISE class are excluded. Curves show the weighted median of the observed flux densities after transformation to the rest frame with $z_{\rm spec}$ and normalization by the fitted total SED at $1\um$; shaded regions span the weighted 16th--84th percentiles. Each covered source has equal total weight within a wavelength bin, and bins with fewer than five sources are masked. The legend reports the number of sources entering each stack, the gray vertical line marks the normalization wavelength, and dotted lines identify selected spectral features.}
    \label{fig:app_confusion_spherex_stack}
\end{figure*}

Figure~\ref{fig:app_confusion_spherex_stack} clearly shows that the confusion is associated with continuous spectral variation. In NEP, the FP stack lies between the TN and TP stacks over much of the near-infrared, with stronger ${\rm Pa}\alpha$ emission and more flux beyond approximately $2\um$ than the correctly classified galaxies. The FN stack instead closely resembles the correctly classified galaxy stack, with little visible ${\rm Pa}\alpha$ enhancement and comparatively weak continuum emission beyond approximately $2\um$. These average differences support the interpretation that some apparent false positives contain low-level activity not captured by the optical spectra, while some false negatives are intrinsically host dominated. In EGS, shallower SPHEREx sampling weakens the coefficient constraints and thereby degrades classification performance, leaving the four NLAGN/other--Galaxy subsets less clearly separated than in NEP. Given the substantial overlap and the selection of these subsamples by classifier outcome, we interpret the stacks only as qualitative evidence for the spectral origin of the confusion, not as a basis for revising individual labels or estimating the catalog-wide label-error rate.